\documentclass[twocolumn,pra,showpacs,multicol,amssymb]{revtex4}
\usepackage{graphicx}
\usepackage{dcolumn}
\usepackage{bm}
\usepackage{graphics}
\usepackage{epsfig,color}
\usepackage{tabu}
\usepackage{array}

\newcommand{\be}{\begin{equation}}
\newcommand{\ee}{\end{equation}}
\newcommand{\bea}{\begin{eqnarray}}
\newcommand{\eea}{\end{eqnarray}}
\newcommand{\la}{\langle}
\newcommand{\ra}{\rangle}

\DeclareUnicodeCharacter{2212}{-}

\begin{document}
\title{Magnetic phase diagram of a spin-1/2 XXZ chain with modulated Dzyaloshinskii-Moriya interaction} 
\author{G. I. Japaridze}
\affiliation{ Ilia State University, Center for Condened Matter Theory and Quantum Computations}
\affiliation{Andronikashvili Institute of Physics, Tbilisi, Georgia}
\author{Hadi Cheraghi and Saeed Mahdavifar}

\affiliation{ $^{1}$Department of Physics, University of Guilan, 41335-1914, Rasht, Iran}

\date{\today}

\begin{abstract}

We consider the ground-state phase diagram of a one-dimensional spin-1/2 XXZ chain with spatially modulated Dzyaloshinskii-Moriya interaction in the presence of applied along with the $\hat{z}$ axis alternating magnetic field. The model is studied using the continuum-limit bosonization approach and the finite system exact numerical technique.  In the absence of the magnetic field, the ground-state phase diagram of the model includes besides the ferromagnetic and gapless Luttiger-Liquid (LL) phases two gapped phases, the composite (C1) phase characterized by the coexistence of the long-range-ordered (LRO) alternating dimerization and the spin chirality patterns, and the composite (C2) phase characterized in addition to the coexisting spin dimerization and alternating chirality patterns, by the presence of LRO antiferromagnetic order. In the case of two-letter gapped phases, in the case of a uniform magnetic field, the commensurate-incommensurate type quantum phase transitions (QPT) from a gapful phase into the gapless phase have been identified and described using the bosonization treatment and finite chain exact diagonalizations studies.
The upper critical magnetic field corresponding to the transition into a fully polarized state has been also determined. It has been shown that the very presence of the staggered component of the magnetic field vapes out the composite (C1) in favour of the composite gapped (C2) phase.

\end{abstract}
\maketitle


\section{Introduction}\label{sec1} 

The study of the quantum phase transitions (QPTs) [{\color{blue}\onlinecite{1}}] in low-dimensional systems has attracted much of current interest particularly in the context of applications in modern quantum information technologies [{\color{blue}\onlinecite{Wen_19}}]. The one-dimensional quantum spin systems, [{\color{blue}\onlinecite{GNT_book_98,Giamarchi_book_04,Takahashi_Book_99,Cabra_Pujol_04,Vasiliev_Volkova_18}}]
because of rich quantum features and complex ground state phase diagrams,  serve as a fascinating playground to investigate QPTs. Although from the rigid point of view the phase transitions occur only in the thermodynamic limit i.e. at an infinite number of system constituent particles, extreme miniaturization of quantum devices makes it important to search for tools, which allow  tracing the very presence of such a transition in the case of finite clusters.  Achieved during the last years substantial progress in ability to control matter at the quantum level, realized on various experimental platforms, including optically trapped ultra-cold atoms [{\color{blue}\onlinecite{ExPlat_UCA}}], photonics [{\color{blue}\onlinecite{ExPlat_7}}], superconducting quantum circuits [{\color{blue}\onlinecite{ExPlat_8}}] and spin chains [{\color{blue}\onlinecite{{ExPlat_9}}}], has triggered wide activity in the use of finite quantum clusters as simulators of system behavior in the thermodynamic limit. 

Already the seminal studies unambiguously confirm that finite quantum spin, fermion or boson system may exhibit all the hallmarks of the QPTs [{\color{blue}\onlinecite{Ent_97,Ent_02,QS_Chains_03,QS_Chains_03a,QS_Chains_03b}}]. The exact calculation of various quantum parameters, such as entanglement [{\color{blue}\onlinecite{Ent_Wang_01,Ent_Solyom_Noack,Ent_SCh_4,Ent_SCh_5,Ent_SCh_6,Ent_SCh_7,Ent_SCh_8}}],  Loschmidt echo [{\color{blue}\onlinecite{Loschmidt_Echo_1,Loschmidt_Echo_2}}], fidelity [{\color{blue}\onlinecite{Fidelity}}] quantum discord (QD) [{\color{blue}\onlinecite{Quant_Discord}}], quantum coherence [{\color{blue}\onlinecite{Quant_Coherence}}], time-dependent quantum behaviour of a complex few spin cluster [{\color{blue}\onlinecite{Grimaudo_etal_19,Grimaudo_etal_20}}] and quantum correlations after sudden quenches [{\color{blue}\onlinecite{Quenches_18a,Quenches_18b,Quenches_20}}]  has been successfully used to identify various QPTs using finite system studies. However, each tool has its limitations for different types of QPTs, and only a proper combination of used markers allows to get a correct description.

The one-dimensional spin systems with Dzyaloshinskii-Moriya interaction (DMI) [{\color{blue}\onlinecite{DMI}}], have become the subject of intensive studies in the last decades, due to their rich quantum nature and high potential of applicability in modern information technologies  [{\color{blue}\onlinecite{Perk_76,Zviagin_89,Alkaraz_90,Aharony_92,Affleck_etal,Aristov_00,Tsvelik_01,YuLu_2003,Derzhko_06,Derzhko_07,
Starykh_etal,Mila_etal_11,Fazio_14,Mahdavifar_08,Mahdavifar_10,Amiri_Langari_13,Mahdavifar_etal_14,Hadi_Saeed_18,Soltani_Saeed_19,
Japaridze_etal_19a,Japaridze_etal_19b,Habibi_etal_19,Lima_19,Thakur_20,Japaridze_etal_20}}].

The Dzyaloshinskii-Moriya interaction, given by the term 
$${\cal H}_{DM} \sim  {\bf D}\cdot \left[{\bf S}_n \times {\bf S}_{m}\right],$$ 
where ${\bf D}$ is an axial vector describes an antisymmetric magnetic exchange between spins located at sites $n$ and $m$. Such an exchange interaction typically arises in non-centrosymmetric bulk magnets [{\color{blue}\onlinecite{Bulk_Systems}}] and  at interfaces between a ferromagnet and an adjacent layer with strong spin-orbit coupling [{\color{blue}\onlinecite{Interfaces}}]. The DMI is responsible for the generation of skyrmions in two-dimensional magnetic structures [{\color{blue}\onlinecite{Skyrmions}}], composite electric and magnetic order in multiferroic materials [{\color{blue}\onlinecite{Multi_Ferroics}}] and for the formation of helical and other unconventional magnetic ordered states in quasi 1D magnetic materials [{\color{blue}\onlinecite{1D_Chains_w_DMI_Exp}}]. 

In real solid-state materials, due to the symmetry restrictions, vector ${\bf D}$ is either uniform or staggered. Respectively, theoretical studies of the spin $S=1/2$ Heisenberg chain and ladder systems with DMI [{\color{blue}\onlinecite{Perk_76,Zviagin_89,Alkaraz_90,Aharony_92,Affleck_etal,Aristov_00,Tsvelik_01,YuLu_2003,Derzhko_06,Derzhko_07,
Starykh_etal,Mila_etal_11,Fazio_14,Mahdavifar_08,Mahdavifar_10,Hadi_Saeed_18,Soltani_Saeed_19}}], exception is only the Ref. [{\color{blue}\onlinecite{Derzhko_07}}] where the $XY$ spin chain with random changes in the sign of DM interactions were studied. 

Recently it has been demonstrated that the DMI can be efficiently tailored with a substantial efficiency factor by an external electric field [{\color{blue}\onlinecite{E-Tailored_DMI_2,E-Tailored_DMI_3,E-Tailored_DMI_4}}].  These achievements open a possibility to manipulate with magnetic properties of the spin system by spatially modulated by the applied external electric potential DMI. In the recent publication, the ground-state phase diagram of a spin-1/2 XXZ Heisenberg chain in the presence of alternating DMI has been studied using continuum-limit bosonization approach [{\color{blue}\onlinecite{Japaridze_etal_19a}}]. It has been shown  that the joint effect of the uniform and the staggered components of the DMI leads to the formation of a gap in the excitation spectrum and, in addition to the standard ferromagnet (FM) and gapless Luttinger-liquid (LL) phase, to the formation of two new unconventional gapped phases in the ground-state: the gapped composite ($C1$) phase characterized by coexistence of the long-range-ordered (LRO) dimerization and the LRO alternating spin chirality patterns and, in the limit of strong exchange anisotropy, of the gapped composite ($C2$) phase characterized in addition to the coexisting spin dimerization and alternating chirality patterns, by the presence of LRO antiferromagnetic order. It has been shown, that the transition from the LL to the $C1$ phase belongs to the Berezinskii-Kosterlitz-Thouless (BKT) universality class, while the transition from $C1$ to $C2$ phase is of the Ising type [{\color{blue}\onlinecite{Japaridze_etal_19a}}]. In the subsequent studies, these QPTs have been also imprinted using quantum correlations studies in finite clusters [{\color{blue}\onlinecite{Japaridze_etal_20}}].

In this paper, we extended our studies of the QPTs in the ground state of the spin-1/2 XXZ chain with alternating DMI and study magnetic phases of the model in the presence of, orientated along with the DM {\bf D} vector, alternating magnetic field. The Hamiltonian under consideration is 
\begin{eqnarray}\label{eq1}
{\cal H} &=&\sum\limits_{n = 1}^{N} \left[J \left( S_n^xS_{n + 1}^x + S_n^yS_{n + 1}^y\right) +J_{z} S_n^zS_{n + 1}^z \right]  \nonumber\\
&+&\sum\limits_{n = 1}^{N}  \left( D_0 + ( - 1)^n D_1 \right)\left( S_n^x S_{n + 1}^y - S_n^y S_{n + 1}^x \right)  \nonumber\\
&-&\sum\limits_{n = 1}^{N}  \left( H + ( - 1)^n H_1 \right) S^z_{n} ,
\end{eqnarray}
where $S_n$ is the spin-1/2 operator on the $n$-th site,  $J>0$ is the exchange coupling, while $D_0$ and $D_1$ 
and $H_0$ and $H_1$ are uniform and staggered parts of the DMI vector and of the magnetic field, respectively.
Studies of the ground-state phase diagram of the XXZ chain in the presence of uniform and staggered magnetic field counts decades and have been revived in many excellent books and papers [{\color{blue}\onlinecite{GNT_book_98,Giamarchi_book_04,Takahashi_Book_99}}].
Therefore below we focus on the study of effects caused by the interplay and competition between the alternating DMI and magnetic field and hallmarks of the corresponding QPTs within the finite chain exact quantum calculations.

The paper is organized as follows: In the forthcoming Sec.~{\color{blue}II}, we consider the exactly solvable limit of the model corresponding to the case $J_{z}=0$. In the Section~{\color{blue}III} the continuum-limit bosonization analysis of the magnetic phase diagram of the model ({\color{blue}\ref{eq1}}) is presented. In the Section~{\color{blue}IV} results of the finite chain, exact diagonalizations studies are presents. Finally, in Section~{\color{blue}V} the summary is put.

\section{\bf The exactly  solvable case}\label{sec:TheModel}

In this section  we consider the exactly solvable case of the model ({\color{blue}\ref{eq1}}) 
at $J_{z}=0$. It is instructive to rewrite the Hamiltonian in the following form
\begin{eqnarray}\label{Hamiltonian_XX_Alt_DM+H}
{\cal H} &=&\sum_{n}\Big[\,\frac{J}{2}
\left(S^{+}_{n}S^{-}_{n+1}+S^{-}_{n}S^{+}_{n+1}\right)\nonumber\\
&+&\frac{i}{2}(D_{0}+(-1)^{n}D_{1})\left(S^{+}_{n}S^{-}_{n+1}
-S^{-}_{n}S^{+}_{n+1}\right)\nonumber\\ 
&-& \left( H + ( - 1)^n H_1 \right) S^z_{n}\,\Big]\, ,
\end{eqnarray}
where $S^{+}_{n}=S^{+}_{x} \pm i S^{+}_{y}$. 
Using the Jordan-Wigner transformations [{\color{blue}\onlinecite{JW_1928}}]
\begin{eqnarray} \label{eq3}
S_{n}^{+} &=& a^{\dagger}_{n} \exp\left(i\pi\sum_{m < n}
a^{\dagger}_{m}a^{\phantom{\dagger}}_{m}\right)\, ,\nonumber\\
\quad S_{n}^{-} &=&
\exp\left(-i\pi\sum_{m < n}a^{\dagger}_{m}a^{\phantom{\dagger}}_{m}\right)a_{n}\, ,\nonumber\\
S^{z}_{n} &=& a^{\dagger}_{n}a^{\phantom{\dagger}}_{n} -
1/2,\label{jordanwigner} \end{eqnarray}
where $a^{\dagger}_{n}$ ($a^{\phantom{\dagger}}_{n}$) is a spinless
fermion creation (annihilation) operator on site $n$, we rewrite the
initial lattice spin Hamiltonian ({\color{blue}\ref{Hamiltonian_XX_Alt_DM+H}}) in
terms of interacting spinless fermions in the following way:
\begin{eqnarray}
\label{Hamiltonian_XX_Alt_DM+H_SF}{\cal H} &=&
\sum_{n}\Big[\frac{J}{2}\left(a^{\dagger}_{n}a^{\phantom{\dagger}}_{n+1}+
a^{\dagger}_{n+1}a^{\phantom{\dagger}}_{n}\right)\nonumber\\
&&\hspace{-5mm}+\frac{i}{2}\left(D_0 + ( - 1)^n D_1 \right)\left(a^{\dagger}_{n}a^{\phantom{\dagger}}_{n+1}-
a^{\dagger}_{n+1}a^{\phantom{\dagger}}_{n}\right)\nonumber\\
&-&\left( H + (-1)^n H_{1} \right)(a^{\dagger}_{n}a^{\phantom{\dagger}}_{n}-1/2)\Big]\, .
\end{eqnarray}
\subsection{The spectrum}

The Hamiltonian ({\color{blue}\ref{Hamiltonian_XX_Alt_DM+H_SF}}) can be easily diagonalized in the momentum space.
Indeed, performing the Fourier transform $a_n =\frac{1}{\sqrt{N}}\sum_k a_k e^{ikn}$ we obtain
\begin{eqnarray}\label{H-JW_Fermions-MS}
{\cal H} &=&\sum_{k} \left[\left(\epsilon(k)-H\right)
\,a^{\dagger}_{k}a^{\phantom{\dagger}}_{k} + \Delta(k)
a^{\dagger}_{k}a^{\phantom{\dagger}}_{ k+\pi}\right]\,
, \end{eqnarray}
where
\begin{eqnarray}
\epsilon(k)&=& J\cos k -D_{0}\sin k, \\
\Delta(k)&=& iD_{1}\cos k \label{Delta-k}+H_{1}\, .
\end{eqnarray}
Diagonalization of the Hamiltonian
({\color{blue}\ref{H-JW_Fermions-MS}}) is straightforward. It is convenient
to restrict momenta within the reduced Brillouin zone $-\pi/2 < k
\leq \pi/2$ and to introduce a new notation $a_{k+\pi}=b_{k}$. In
these terms the Hamiltonian reads
\begin{eqnarray}
{\cal H} &=&\sum_{k}{}^{\prime} \,\Big[\left(\epsilon(k)-H\right)a_{k}^{\dag}
a^{\phantom{\dagger}}_{k}
- \left(\epsilon(k)+ H\right)b_{k}^{\dag}b^{\phantom{\dagger}}_{k}\nonumber\\
&&\hspace{10mm}+\Delta(k)\, a_{k}^{\dag}b^{\phantom{\dagger}}_{k} + \Delta^{\ast}\,(k)b_{k}^{\dag}a^{\phantom{\dagger}}_{k} \Big]\, ,
\end{eqnarray}
where prime in the sum means that the summation is taken over the
reduced Brillouin zone $-\pi/2 < k \leq \pi/2$. Using the unitary
transformation
\begin{eqnarray} \label{eq9}
a_{k} &=& \cos\phi_{k}\,\alpha_{k} +\sin\phi_{k}e^{i\theta_{k}}\,\beta_{k}\,b , \nonumber\\
b_{k} &=& -\sin\phi_{k}e^{-i\theta_{k}}\,\alpha_{k} +\cos\phi_{k}\,\beta_{k}\, .
\end{eqnarray}
and choosing
\begin{eqnarray}
\tan\,\theta_{k}&=&D_{1}\cos(k)/\epsilon(k)\, ,\nonumber\\  
\tan\,2\phi_{k}&=& \sqrt{\epsilon^{2}(k) + D_{1}^{2}\cos^{2} k} /H_{1}\, ,  
\end{eqnarray}
we finally obtain
\begin{eqnarray}
{\cal H}
&=&\sum_{k}{}^{\prime}\, \Big[\,E_{+}(k)\alpha_{k}^{\dag}\alpha^{\phantom{\dagger}}_{k}
+ E_{-}(k)\beta_{k}^{\dag}\beta^{\phantom{\dagger}}_{k}\,\Big]\, ,
\end{eqnarray}
where 
\bea\label{EigenValues}
E_{\pm}(k)&=&-H \pm \sqrt{\epsilon^{2}(k) + D_{1}^{2}\cos^{2} k + H_{1}^{2}} \, .
\eea
The minimum of the upper, "$\alpha$" band,  and the maximum of the lower, "$\beta$" band, is reached at 
\begin{eqnarray}\label{eq8}
k^{\ast}=\pi/2+k_{0}  =\frac{\pi}{2}-\frac{1}{2}\tan^{-1} \left(\frac{2J D_0}{J^{2}-D_{0}^{2}+D_{1}^{2}}\right)\, .
\end{eqnarray}
At this point the dispersion relation ({\color{blue}\ref{EigenValues}}) shows a gap  
\begin{eqnarray}\label{Gap}
E_{+}(k^{\ast})- E_{-}(k^{\ast})=2\Delta_{0}\, ,
\end{eqnarray}
where
\begin{eqnarray}\label{Gap_XX_D1}
&&\hspace{-5mm}\Delta_{0}=\sqrt{H^{2}_{1}+\frac{1}{2}\left(J^{*2}-\sqrt{J^{*4} -(2D_{0}D_{1})^2}\right)}\nonumber\\
&&\simeq \sqrt{H^{2}_{1}+\frac{D^{2}_{0}D^{2}_{1}}{J^{*2}}}\, , \quad\, {\mbox at}\quad \,\, D_{0}D_{1}\ll J^{*2}
\end{eqnarray}
and $J^{*}=\sqrt{J^{2}+D^{2}_{0}+D_{1}^{2}}$.

Respectively the maximum of the upper $\alpha$ band and the minimum of the lower "$\beta$" band is reached at $k_{0}$ and are equal to
\begin{eqnarray}\label{Gap_XX_D1}
E_{\pm}(k_{0})&=&-H \pm \sqrt{ \Delta_{0}^{2}+\sqrt{J^{*4}-(2D_{0}D_{1})^2}}\, ,
\end{eqnarray}

\begin{figure}[t]
\centerline{\psfig{file=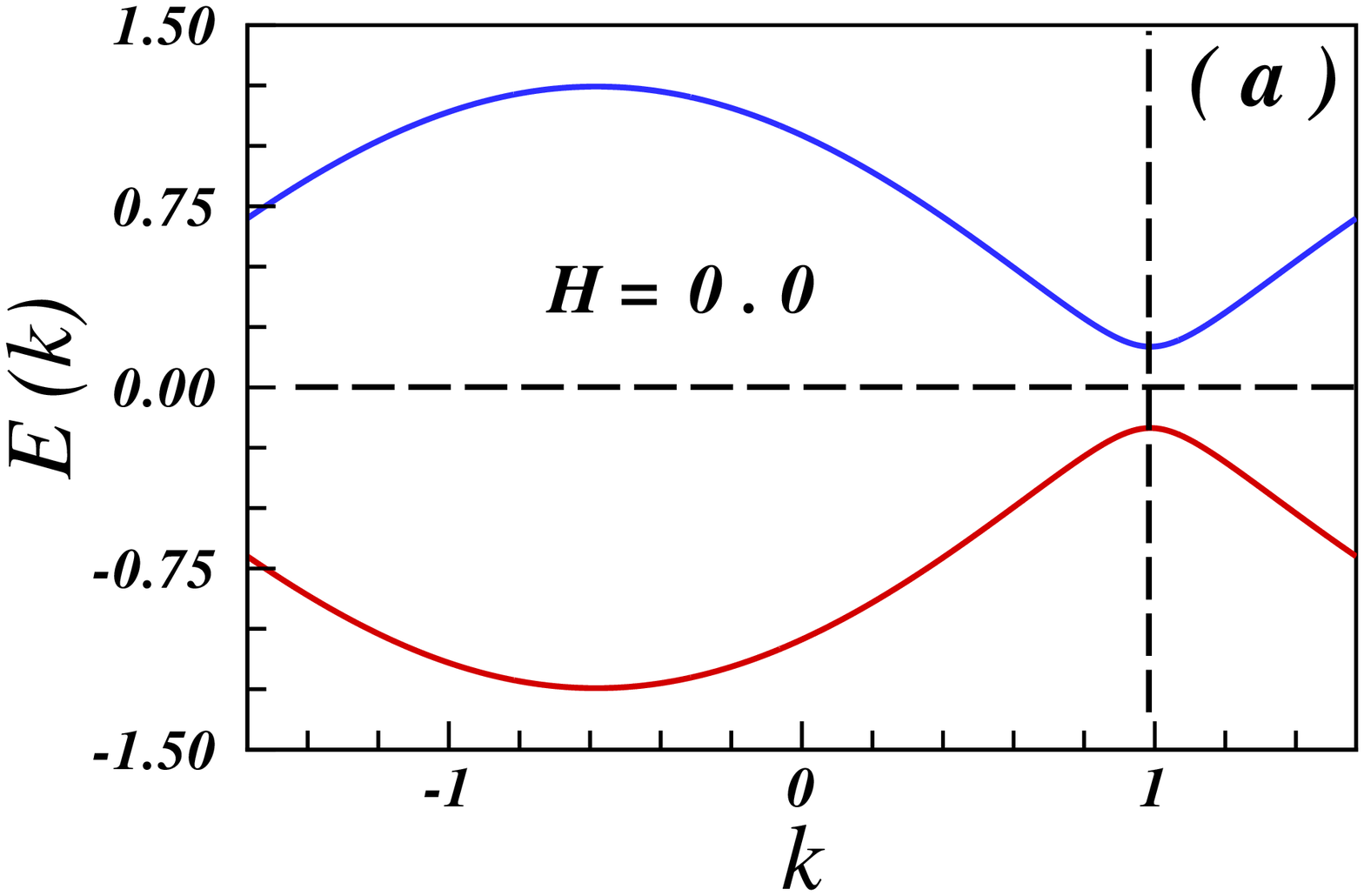,width=1.7in} \psfig{file=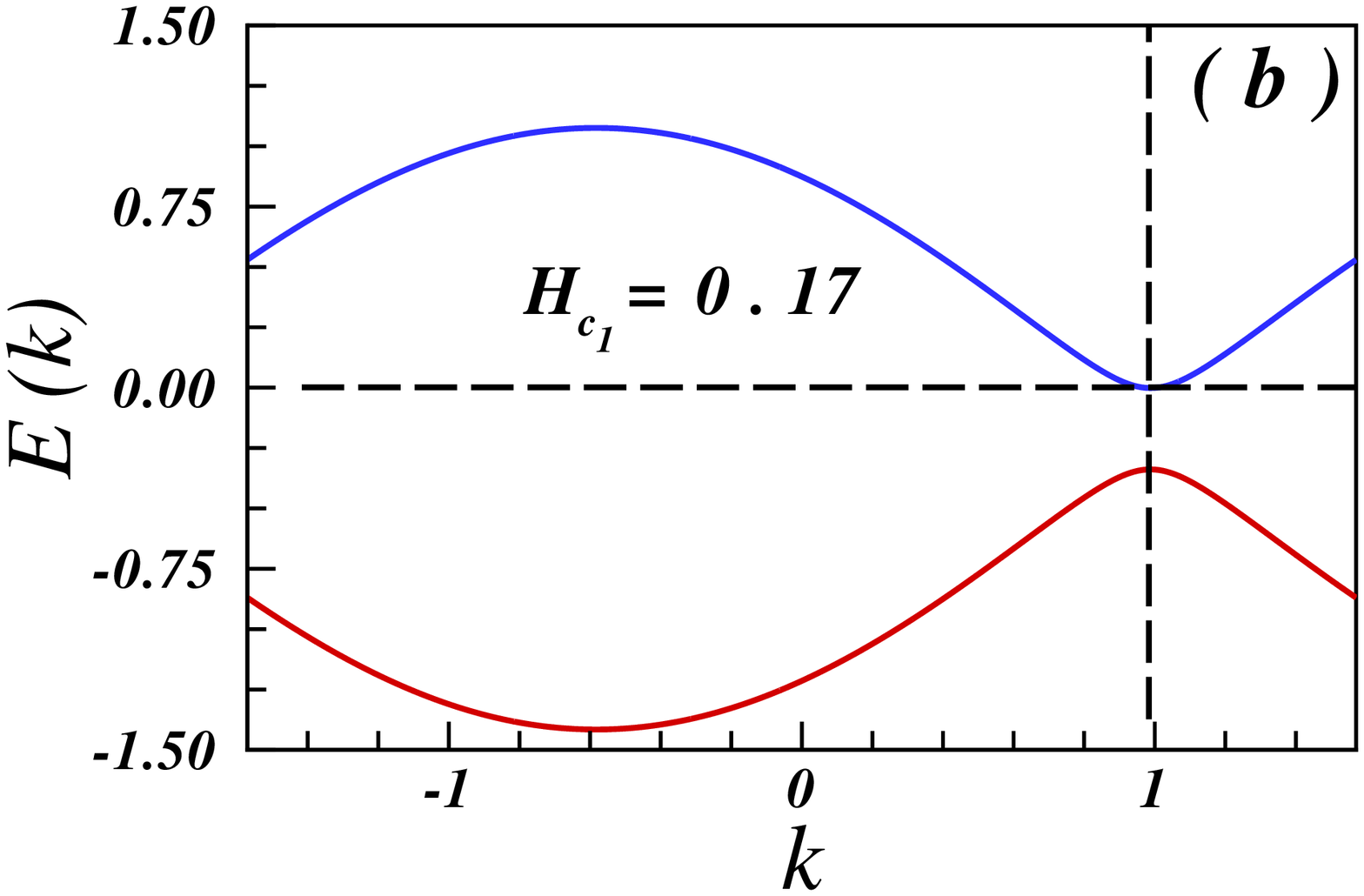,width=1.7in}}
\centerline{\psfig{file=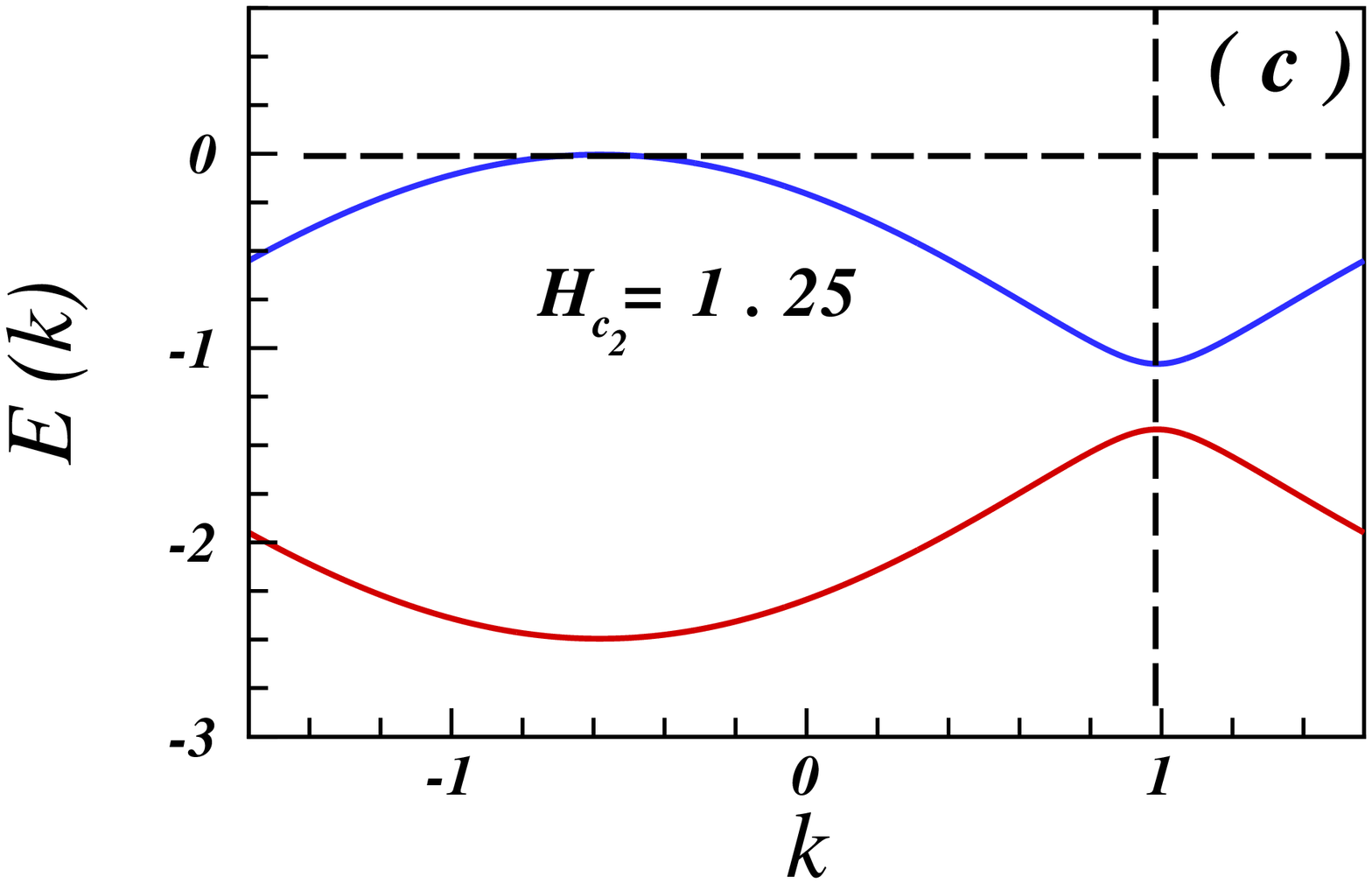,width=1.7in} \psfig{file=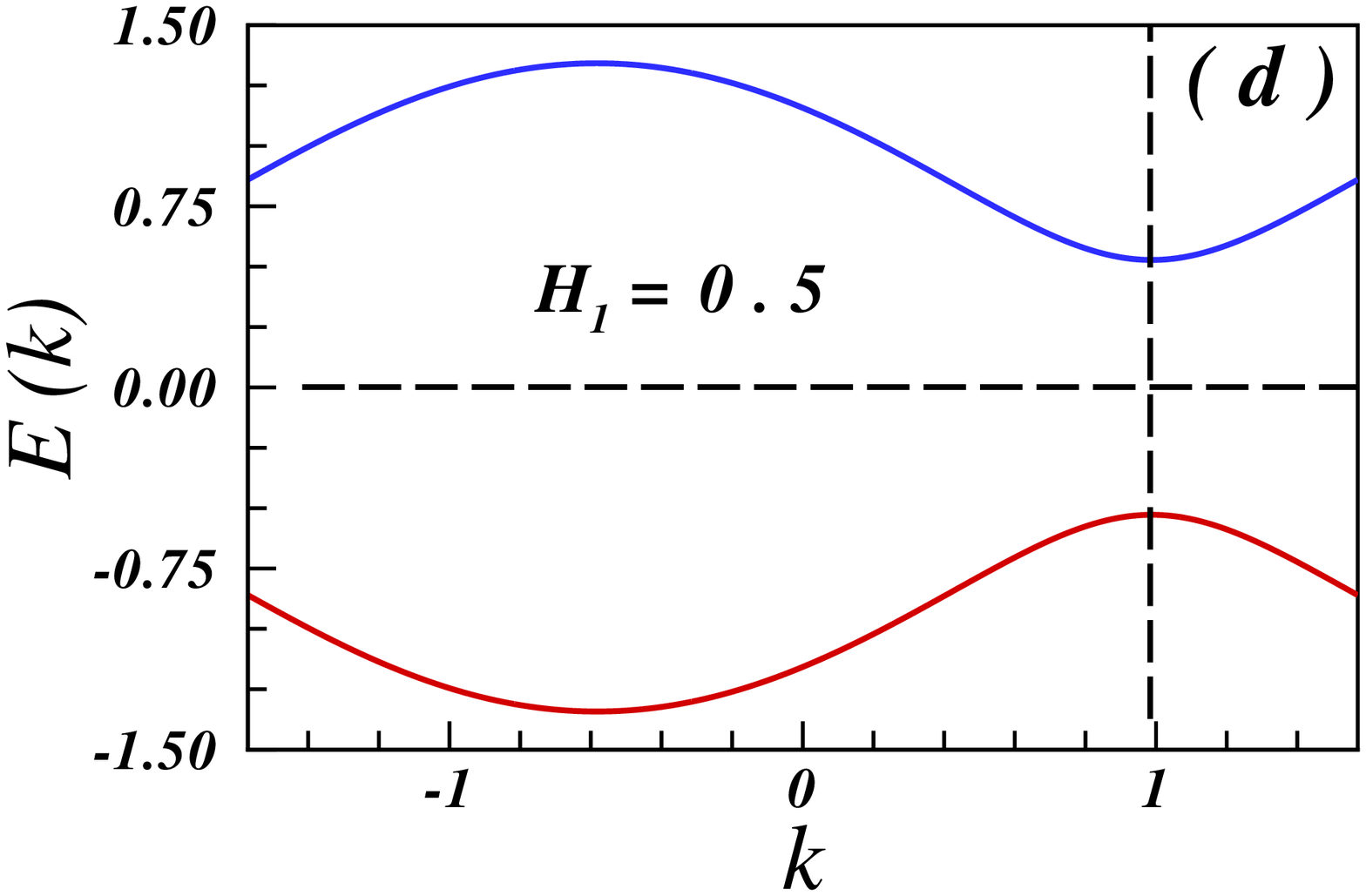,width=1.7in}}
\caption{(color online) Single particle dispersion relations  for a DMI as $D_0=0.7$ and $D_1=0.3$ for the  uniform magnetic fields  as (a) $H=0.0$, (b) $H_{c_1}=0.17$,  (c) $H_{c_2}=1.25$, and also for the case of the staggered  magnetic field as (d) $H_1=0.5$. The horizontal and vertical black dashed lines hint $E(k)=0.0$ and $k^*$, respectively. }
\label{energy-u}
\end{figure}

As it follows from ({\color{blue}\ref{Gap_XX_D1}}), in the absence of the magnetic field, only synergic action of the uniform nor staggered components of the DMI provide the emergence of a gap in the excitation spectrum [{\color{blue}\onlinecite{Japaridze_etal_19a}}]. In marked contrast, the effect of uniform and staggered components of the magnetic field are mutually exclusive -- if the staggered component contributes additively to DMI and enlarges the gap, the uniform component serves as a chemical potential and at $H>H_{c_1}=\Delta_{0}$ leads to the gapless excitation spectrum (see Fig.~{\color{blue}\ref{energy-u}}). 

\subsection{The ground-state structure, order parameters, and critical points}

Obtained in the considered exactly solvable case single-particle dispersion relations ({\color{blue}\ref{EigenValues}}) allow to identify key points of the ground state magnetic phase diagram, the character of the low-energy excitation spectrum, and critical points. These results will serve as a milestone for the general, nonintegrable case, considered in the following sections. 

In the ground state all states with negative energy are filled and with positive energy  are empty. Since $E_{-}(k)<0$ for arbitrary $H$ all states in the lower band are completely filled and $n_{\beta}(k)=\langle 0|\beta^{\dag}_k \beta_k |0\rangle=1$ for $k \in [-\pi/2,\pi/2]$. 

At $H<H_{c_1}$ the upper band is empty $n_{\alpha}(k)=\langle 0|\alpha^{\dag}_{k} \alpha_{k}|0\rangle=0$. However,  when the uniform component of the magnetic field exceeds the critical value, at $H>H_{	c_1}$ states of the upper band between two Fermi points
\begin{eqnarray}
k_F^{\pm}  = k_0 \pm  \arcsin \sqrt{\frac{H^{2} - \Delta_{0}^{2}}{\sqrt{ J^{*4}-(2D_0 D_1)^2}}},
\end{eqnarray}
become occupied, and at 
\begin{eqnarray}\label{Hc1}
H\geq H_{c_2}=\sqrt{\Delta_{0}^{2}+\sqrt{ J^{*4}-(2D_0 D_1)^2}}\, 
\end{eqnarray}
all states in the upper band are also occupied. Thus 
\begin{eqnarray}\label{eq24}
n_{\alpha}(k)= \left\{ {\begin{array}{*{15}{c}}
{\begin{array}{*{20}{c}}
0 \quad {\mbox for} \quad  k \in [-\pi/2,\pi/2] \quad {\mbox at}\quad  {H < {H_{c_1}}}
\end{array}}\\
{\begin{array}{*{20}{c}}
\hspace{-11mm}1  \quad {\mbox for} \quad  k \in \Lambda(k) \quad {\mbox at}\quad { H > H_{c_1}}
\end{array}}
\end{array}} \right.
\end{eqnarray}
where 
\begin{eqnarray} \label{eq25}
\Lambda (k) = \left\{ {\begin{array}{*{20}{c}}
{\begin{array}{*{20}{c}}
{[k_F^ - ,k_F^ + ]}\\
{[ - \pi /2,\pi /2]}
\end{array}}&{\begin{array}{*{20}{c}}
{H_{c_1} < H < H_{c_2}}\\
{H > H_{c_2}}
\end{array}}
\end{array}} \right.
\end{eqnarray}   

\begin{figure}[t]
\centerline{\psfig{file=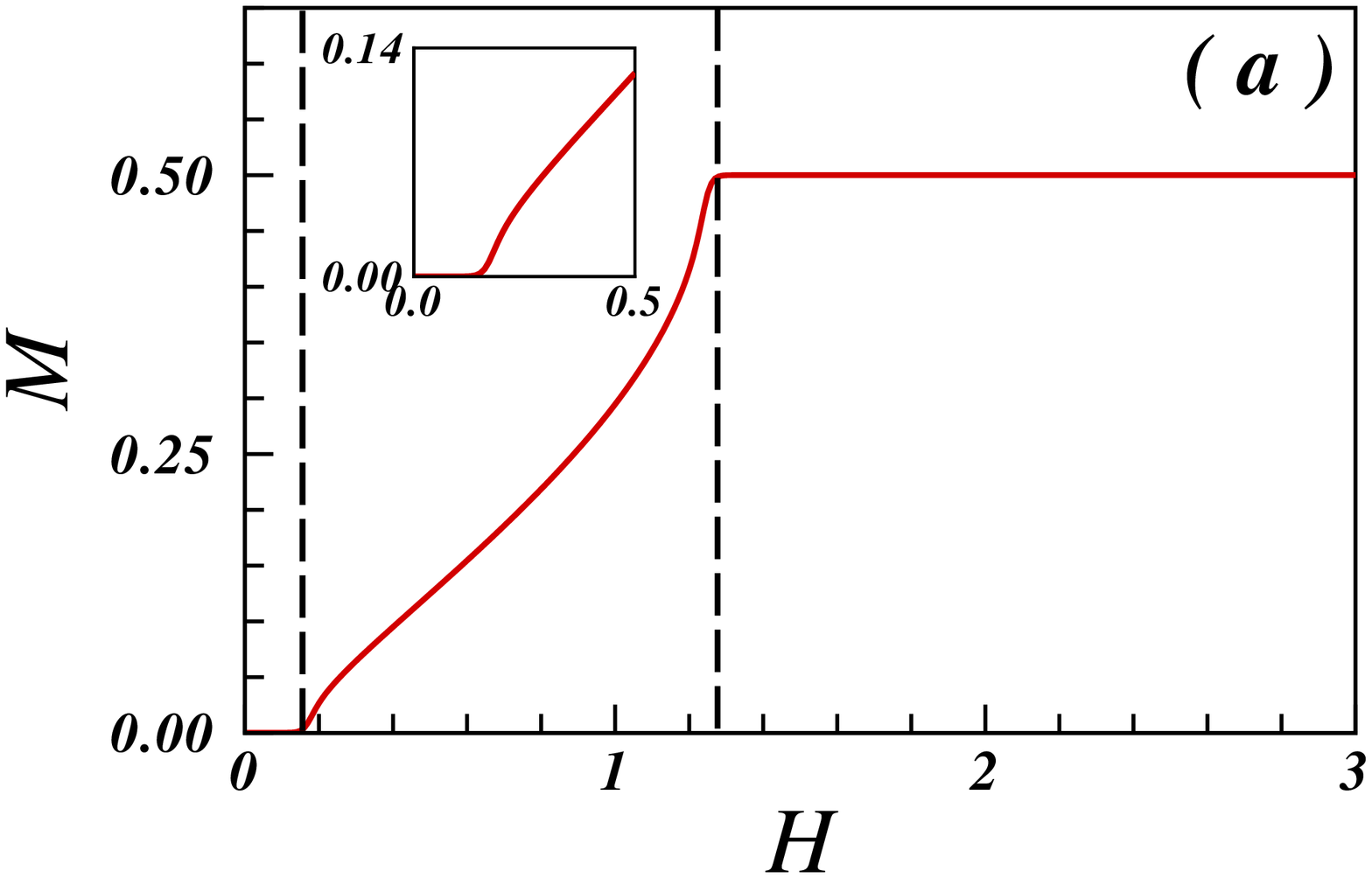,width=1.7in} \psfig{file=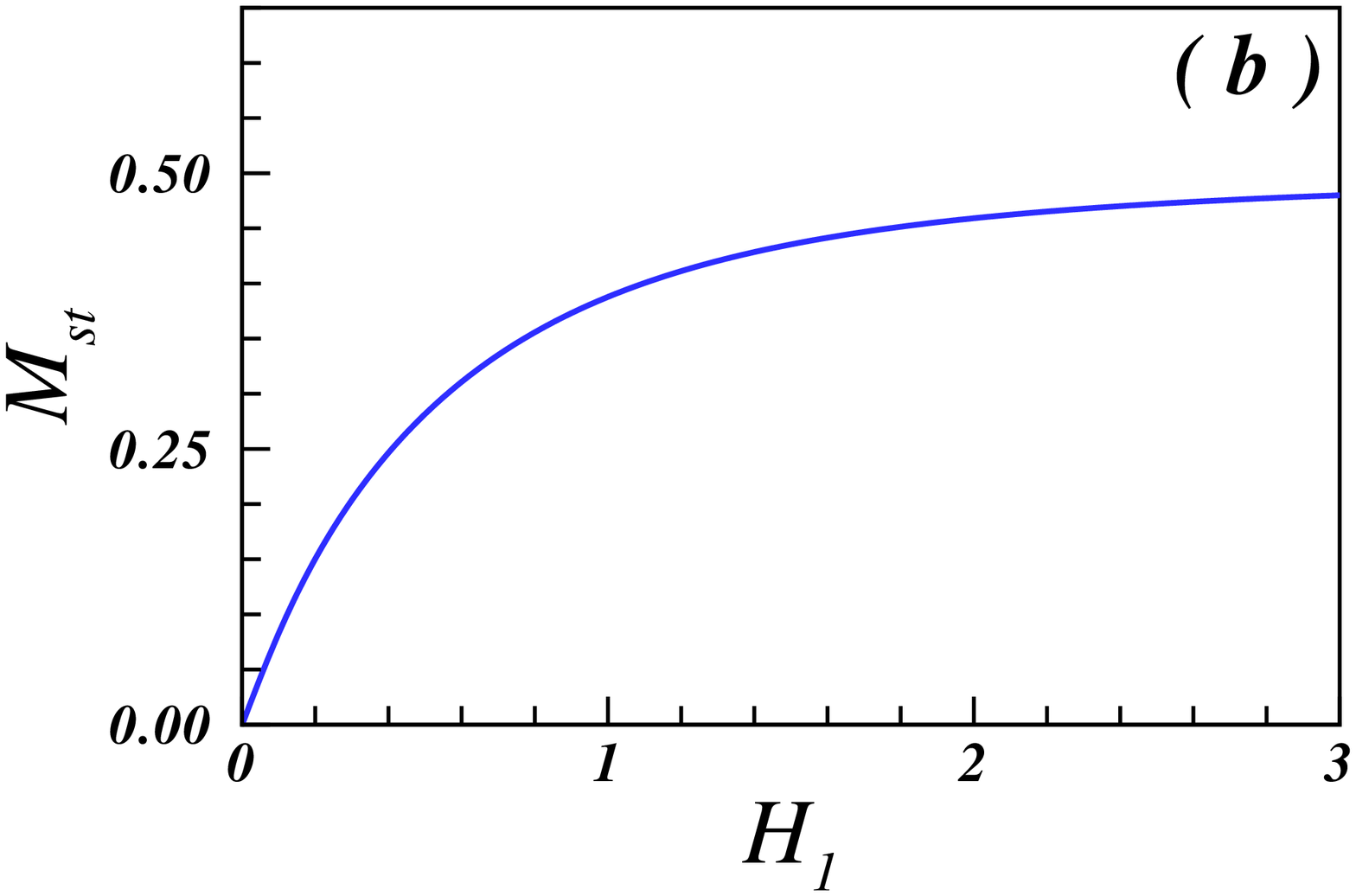,width=1.7in}}
\caption{(color online) (a) Magnetization and (b) staggered magnetization versus the  magnetic fields  for a given DMI as $D_0=0.7$ and  $D_1=0.3$. The inset in (a) expresses a square-root behavior of the magnetization near to $H_{c_1}$. The vertical black dashed lines in (a) point out the critical uniform  ones. }
\label{Mag}
\end{figure}

Using ({\color{blue}\ref{eq24}}) one easily obtains that at $H<H_{c_1}$ net magnetization of the system is zero
\begin{eqnarray}
&&\hspace{-5mm}M = \frac{1}{N}\sum_{n}\la 0 |S_{n}^{z}|0 \ra=\frac{1}{2\pi}\int_{-\pi/2}^{\pi/2}dk\, n_{\beta}(k)
-\frac{1}{2}=0\, ,\nonumber
\end{eqnarray}
while, for $H > H_{c_1}$ the system acquires a finite magnetization per site
\begin{eqnarray}
&&M=\frac{1}{2\pi}\int_{-k_F^{-}}^{k_F^{+}}\,dk\, n_{\alpha}(k)=\frac{1}{2\pi}\left(k_{F}^{+}-k_{F}^{-}\right)\nonumber\\
&&\hspace{5mm}=\frac{1}{\pi} \arcsin \sqrt{\frac{H^{2} - \Delta_{0}^{2}}{\sqrt{ J^{*4}-(2D_0 D_1)^2}}}\, .
\end{eqnarray}
In the close proximity to the transition point, at $H > H_{c_1}=\Delta_{0}$ but $H - H_{c_1} \ll H_{c_1}$ the magnetization 
shows a square-root behavior
\begin{eqnarray}
&&M \simeq {\cal C} \left(H - H_{c_1}\right)^{1/2}\, ,
\end{eqnarray}
while the magnetic susceptibility is divergent
\begin{eqnarray}
&&\chi \simeq \frac{1}{2}{\cal C}\left(H - H_{c_1}\right)^{-1/2}\, ,
\end{eqnarray}
where 
$${\cal C}=\frac{1}{\pi}\sqrt{\frac{2\Delta_{0}}{\sqrt{J^{*4}-(2D_0 D_1)^2)}}}\, .$$

At $H > H_{c_1}$ the excitation spectrum is gapless. As displayed in Fig.~{\color{blue}\ref{Mag}(a)}, as soon as the uniform magnetic field exceeds the first critical field, the magnetization starts to boost and exhibits a monotonic behavior up to the second critical field $H_{c_2}$ where the magnetization reaches its saturation value $M=1/2$.
It is explicit at a certain magnetic field as $H_{c_1}<H_T<H_{c_2}$, the concavity of the graph of the magnetization changes.  The insert in Fig.~{\color{blue}\ref{Mag}(a)} clearly shows the square-root behavior in the vicinity of $H_{c_1}$.
On the other side, Fig.~{\color{blue}\ref{Mag}(b)} exhibits that right after the exert of the staggered case, the Neel order induces. In such a situation enhancing the staggered field leads to the increment of the staggered magnetization, thereby at $H_1 \to \infty $  it tends to the value of 1/2 albeit the saturation is never achieved for any finite staggered field.

\begin{figure}[t]
\centerline{\psfig{file=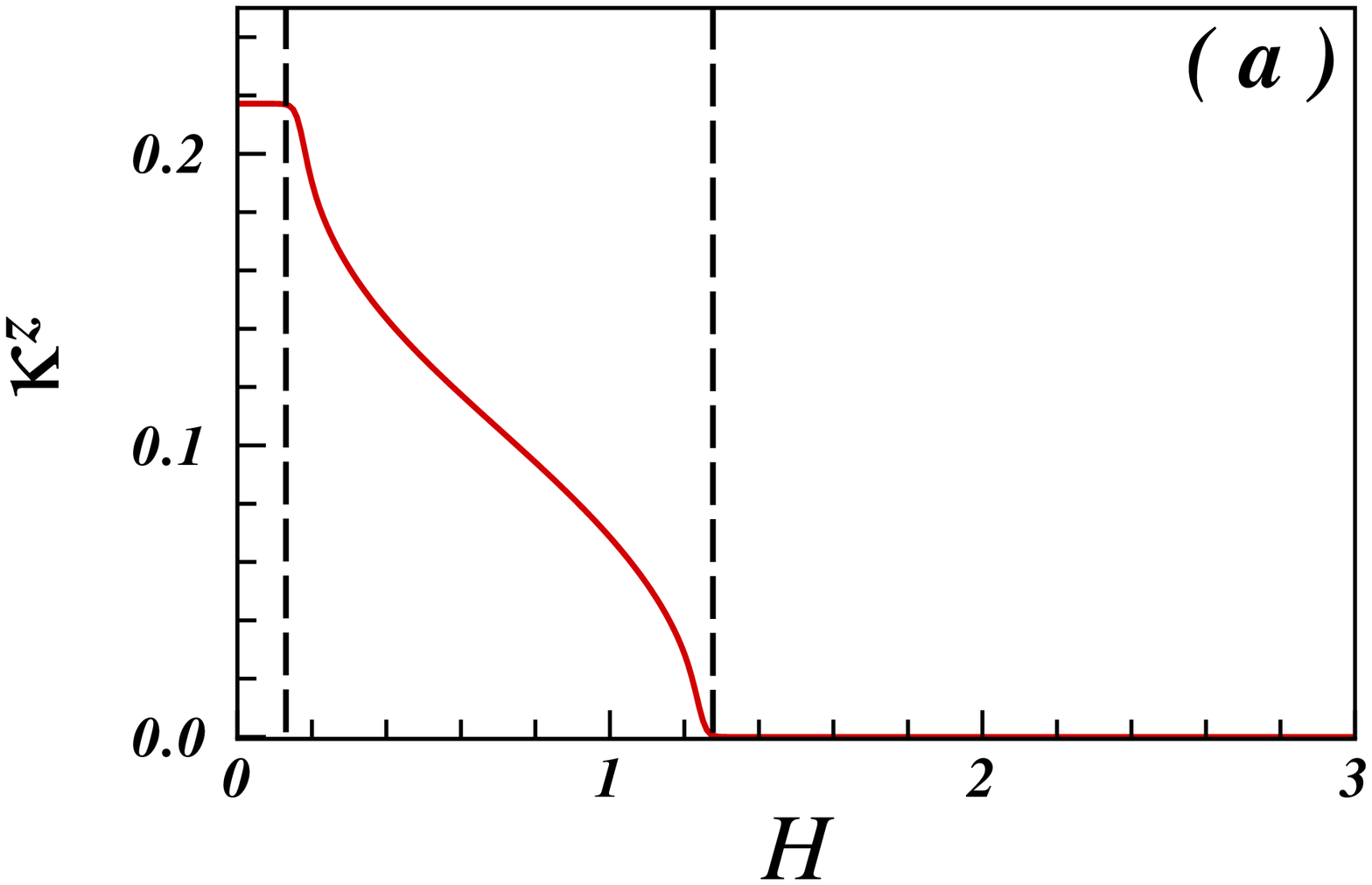,width=1.7in} \psfig{file=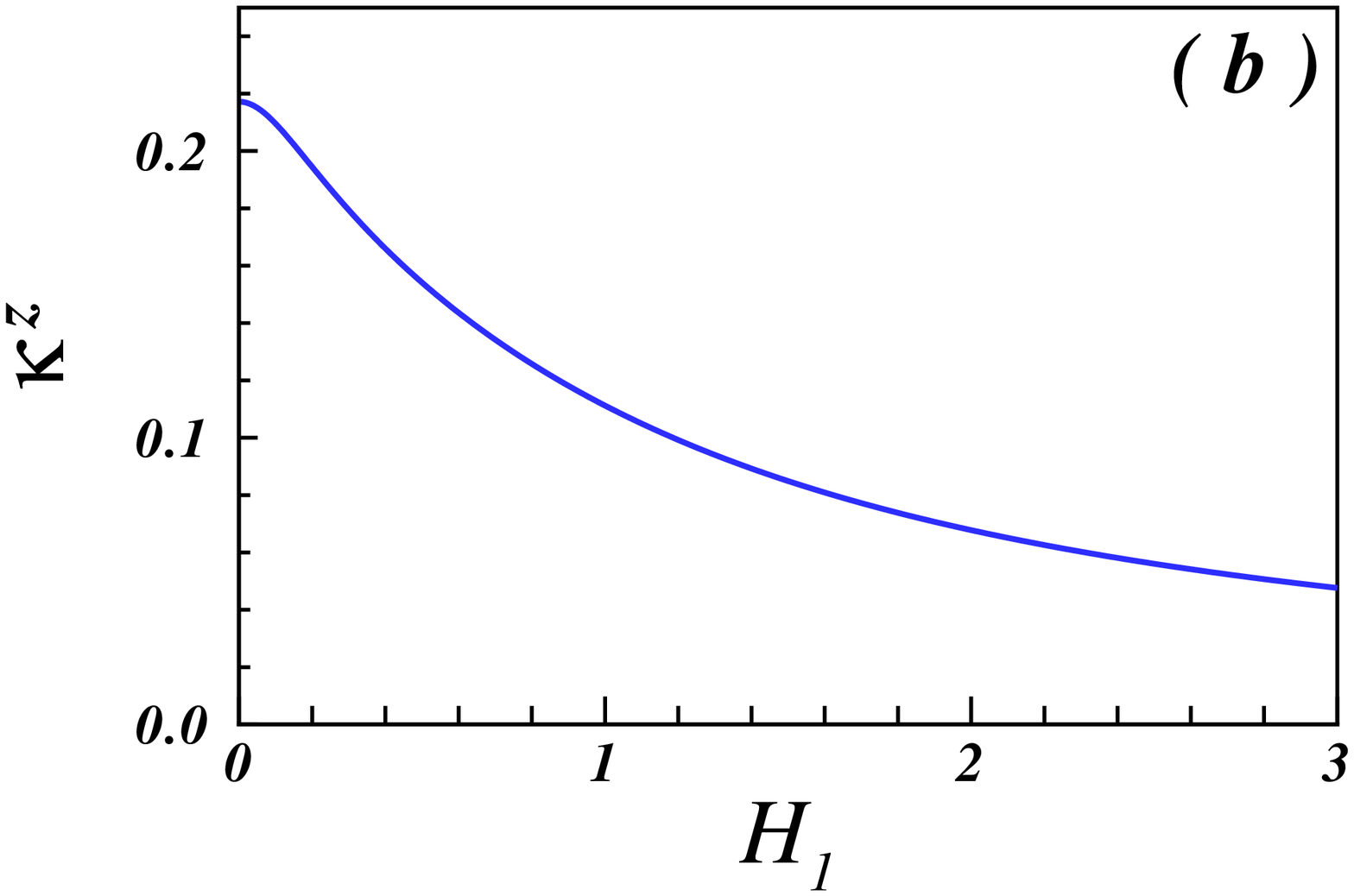,width=1.7in}}
\caption{(color online)  The alternating chiral order parameter for the DMI  values $D_0=0.7$ and $D_1=0.3$ versus (a) the uniform and (b) the staggered magnetic fields. }
\label{Chi}
\end{figure}

Order parameters always serve as an excellent tools to identify QPT-s  [{\color{blue}\onlinecite{1}}] even in the case of  nonequilibrium criticality [{\color{blue}\onlinecite{48}}]. At $H<H_{c_1}$ the system is in the gapped phase, with a fixed LRO structure of correlations in the ground state.  To identify this structure consider the following set of order parameters:

a) the staggered magnetization
\begin{eqnarray}
M_{st} & = & \frac{1}{N}\sum_{n}(-1)^{n}\la 0|S_{n}^{z}|0\ra \nonumber\\
& = & \frac{1}{2N}\sum\limits_{k=-\pi/2}^{\pi/2} \cos (2{\phi _k}), \, 
\end{eqnarray}

b) the staggered transverse spin dimerization on the plane vertical over the $z$-axis
\begin{eqnarray}
{\cal D}^{\perp} &=& \frac{1}{N}\sum\limits_{n = 1}^{N} (-1)^{n} {\left\langle S_n^x S_{n+1}^x+ S_n^y S_{n+1}^y\right\rangle }   \nonumber\\
& =&\frac{1}{2N}\sum\limits_{k\rq{}} \sin(2\phi_{k\rq{}}) \omega (\theta _{k\rq{}},k\rq{})\,  
\end{eqnarray}
where  $\omega (\theta _{k\rq{}},k\rq{}) = \cos (\theta _{k\rq{}})- \cos (k\rq{} - \theta _{k\rq{}})$ and 

c) the staggered transverse spin chirality order parameter [{\color{blue}\onlinecite{Chubukov_91}}] 
\begin{eqnarray}
{\cal K}^z &=& \frac{1}{N}\sum\limits_{n = 1}^{N} (-1)^{n} {\left\langle S_n^x S_{n+1}^y- S_n^y S_{n+1}^x\right\rangle } \nonumber\\
& =&\frac{1}{2N}\sum\limits_{k\rq{}} \sin(2\phi _{k\rq{}}) \varpi (\theta _{k\rq{}},k\rq{})\,.  
\end{eqnarray}
where $\varpi (\theta _{k\rq{}},k\rq{}) =  \sin (\theta _{k\rq{}})-\sin (k\rq{} - \theta _{k\rq{}})$. Note that $k\rq{}$ refers to $[-\pi/2,\pi/2]$ for $H<H_{c_1}$ and $\Lambda (k)$ for  $H>H_{c_1}$. The computing of the order parameters is put in detail in Appendix. 

\begin{figure}[t]
\centerline{\psfig{file=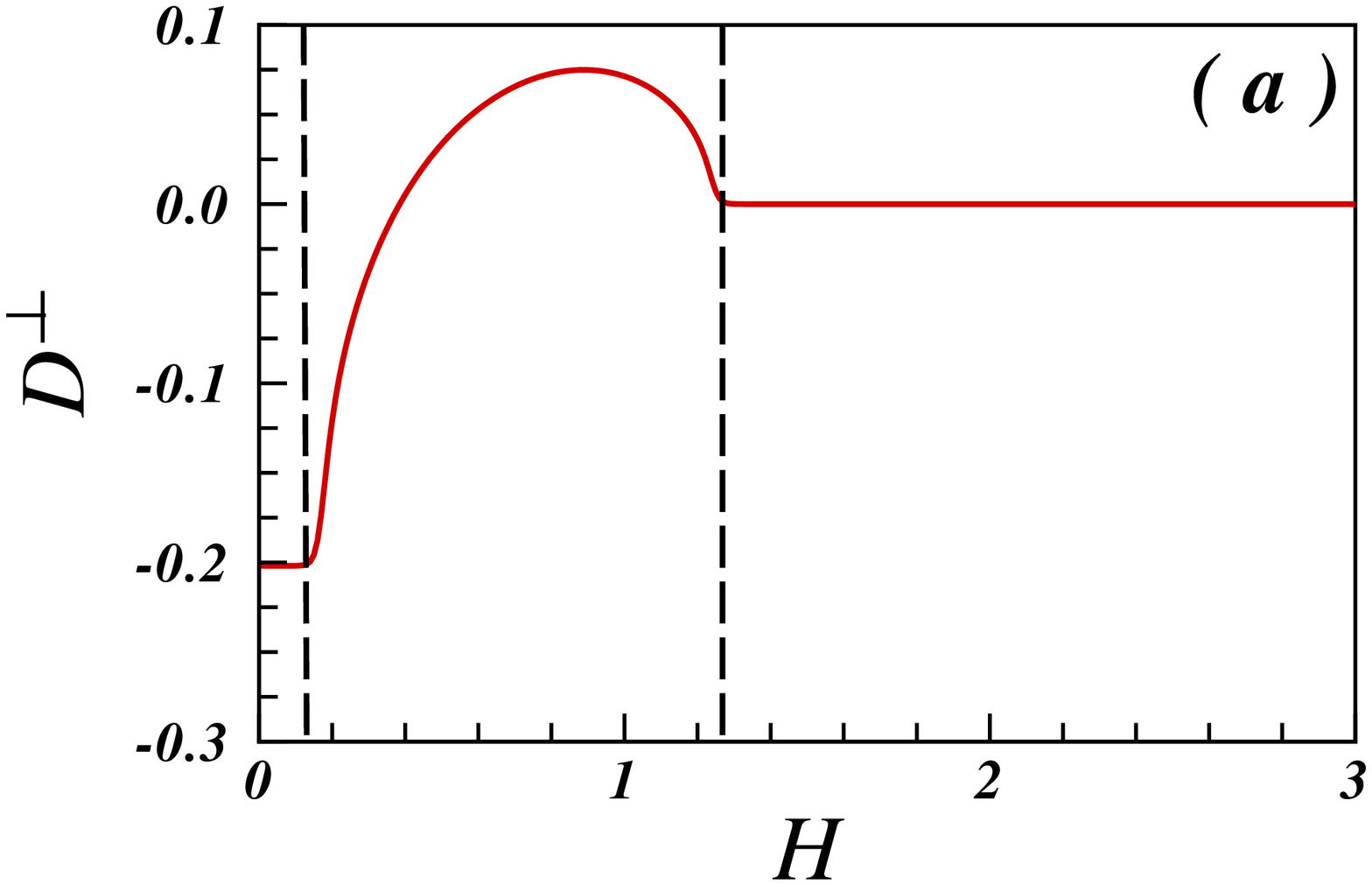,width=1.7in} \psfig{file=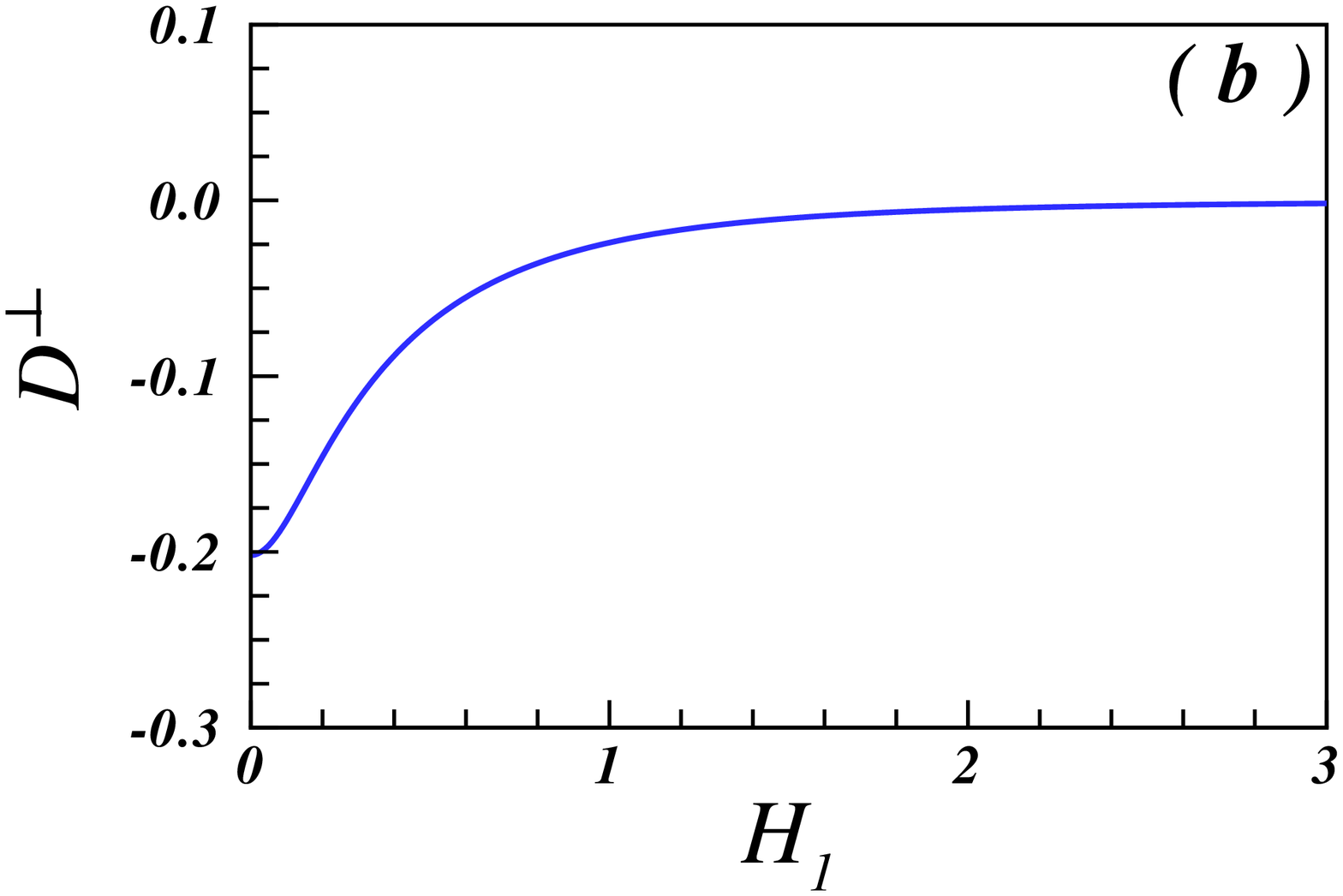,width=1.7in}}
\caption{(color online) (a) The dimer order parameter for  DMI interaction values as $D_0=0.7$ and $D_1=0.3$ as function of (a) the uniform and (b) the staggered magnetic fields. }
\label{Dim}
\end{figure}

In the absence of a magnetic field, the ground state of the model ({\color{blue}\ref{Hamiltonian_XX_Alt_DM+H}}) is characterized by the coexistence of the LRO alternating transverse dimerization and alternating chirality order [{\color{blue}\onlinecite{Japaridze_etal_19a}}].  In Fig.~{\color{blue}\ref{Chi}} and Fig.~{\color{blue}\ref{Dim}} we have plotted the corresponding order parameters as a function of the applied uniform or staggered magnetic field for a chain with DMI values $D_0=0.7,~ D_1=0.3$. The very presence of plateau at $H<H_{c_1}$ in Fig.~{\color{blue}\ref{Chi}(a)} and Fig.~{\color{blue}\ref{Dim}(a)} demonstrate that the system endures in front of the uniform magnetic field. Immediately after the uniform component of the magnetic field rises from  $H_{c_1}$ the chirality and dimerization parameters, {\em which still marks the response of the system on the explicitly broken by the alternating DM translation symmetry, already in the given gapless phase}, decreases monotonically so as to provide the zero values exactly at the saturation critical field $H_{c_2}$. At $H>H_{c_2}$ the system is fully polarized, without any trace of chirality of dimer order. The interesting is at $H_T$, the value of the dimer is zero. On the contrary, in this certain field, the chiral order has a nonzero value. As a consequence, $H_T$ is the field that can distinguish the coexistence of these two orders.
Conversely, in the Fig.~{\color{blue}\ref{Chi}(b)} and Fig.~{\color{blue}\ref{Dim}(b)} the alternating chiral and dimer order parameter are plotted as a function of staggered magnetic field. As it follows from these figures both parameters monotonically decay with increasing $H_{1}$ asymptotically approaching zero at $H_{1} \rightarrow \infty$.

To conclude this section let us review the obtained results. As was extracted, the order parameters accurately respond to change of the magnetic fields; in contrary to the staggered case which just adds Neel order in the system, the uniform field unveils two critical fields, $H_{c_1}$ and $H_{c_2}$. In a range of magnetic fields, we see the conditions where the orders coexist.
Moreover, albeit the staggered field tends the system to have a saturated staggered magnetization but annihilates the other orders.

\section{The continuum-limit bosonization treatment}

To continue our analysis of the model ({\color{blue}\ref{eq1}}) in the whole area $\gamma=J_{z}/J >-1 $, in this Section we present results obtained by the continuum-limit bosonization  treatment. Below we follow the route developed in 
[{\color{blue}\onlinecite{Japaridze_etal_19a}}] to study the ground state magnetic phase diagram of a spin $S=1/2$ $XXZ$ chain with alternating DMI.

\subsection{Gauging away the DM interaction}

We start from the Hamiltonian ({\color{blue}\ref{eq1}}). In analogy with the case of a spin chain with uniform DMI it is instructive to gauge away the alternating DM term by a position-dependent rotation of spins about the ${\hat z}$ axis [{\color{blue}\onlinecite{Perk_76,Japaridze_etal_19a}}]. We introduce new spin variables ${\bf \tau}_{2n}$ and ${\bf \tau}_{2n+1}$ by performing a site-dependent rotation of spins along the chain around the ${\hat z}$ axis with relative angle $\vartheta_{-}$ for spins at consecutive odd-even sites ($2n-1,2n$) and $\vartheta_{+}$ for spins at consecutive even-odd sites ($2n,2n+1$), as
\begin{eqnarray}\label{S2Tau}
  S^{+}_{2n-1}&=& e^{i(n-1)(\vartheta_{-}+\vartheta_{+})} \tau^{+}_{2n-1},\nonumber\\
 S^{+}_{2n} &=&  e^{in\vartheta_{-}+i(n-1)\vartheta_{+}}\tau^{+}_{2n},\\
S^{+}_{2n+1}&=& e^{in(\vartheta_{-}+\vartheta_{+})} \tau^+_{2n+1},\nonumber\\
 S^z_{2n \pm 1}&=& \tau^z_{2n \pm 1} ~~;~~S^z_{2n}= \tau^z_{2n}\, .\nonumber
\end{eqnarray}
Inserting ({\color{blue}\ref{S2Tau}}) in ({\color{blue}\ref{eq1}}) and choosing angles $\vartheta_{\pm}$ such that
$$
\tan\vartheta_{\pm}= D_{\pm}/J,
$$
one cancels the DM coupling and obtains, in terms of new $\tau$ spin variables, the Hamiltonian of $XX$ chain with an alternating exchange in the presence of alternating transverse magnetic field
\begin{eqnarray}\label{XXZ_w_Modulated_DMI_Rotated}
{\cal H}&=& \sum_{n=1}^N \Big[\frac{ \tilde{J}}{2}(1-(-1)^{n}\delta )\left(\tau^{+}_{n}\tau^{-}_{n+1}+ \tau^{-}_{n}\tau^{+}_{n+1}\right)\nonumber \\
&+&J_{z}\tau^{z}_{n}\tau^{z}_{n+1}-(H +(-1)^{n}H_{1})\, \tau^{z}_{n}\Big]\, .
\end{eqnarray}
Here  
\begin{eqnarray}\label{eq26}
\tilde{J}&=&\frac{1}{2}\left(J_{+}+J_{-}\right)\simeq J^{\ast}+{\cal O}\left(D_{i}/J\right)^{4}, \\
\delta \tilde{J}  &=&\frac{1}{2}\left(J_{+}-J_{-}\right)\simeq \frac{D_{0}D_{1}}{J^{\ast 2}}+{\cal O}\left(D_{i}/J\right)^{4},
\end{eqnarray}
at $D_{\pm} \ll J$ ( $i=\pm$) and $J_{\pm} = \sqrt{J^2+D_{\pm}^{2}}$.\\

Thus, after the gauge transformation, we obtain the Hamiltonian of the spin-$1/2$ $XXZ$ Heisenberg chain with alternating transverse exchange [{\color{blue}\onlinecite{Derzhko_07}}] in the presence of an alternating magnetic field. The analytical and numerical studies of the spin-$1/2$ Heisenberg chain with dimerized exchange count decades. The gapped excitation spectrum of bond alternating spin-$1/2$ AFM Heisenberg chain was first predicted by Bulaevskii in 1963 [{\color{blue}\onlinecite{Bulaevskii_63}}]. The analytical and numerical studies of the spin-$1/2$ Heisenberg chain with dimerized exchange include studies of the ordered phases and of the QPT-s in the ground state[{\color{blue}\onlinecite{Cross_Fisher_79,Hida_92,Yamamoto_97,Takayoshi_Sato_10,Qiang_et_al_13,Wang_et_al_13,Giamarchi_etal_18a,Ueda_Onoda_20}}],
of the excitation spectrum [{\color{blue}\onlinecite{Harris_73,Bonner_Bloete_82,BKJ_98,Affleck_Lect_Notes_07}}]
as well as of the magnetic and thermal properties [{\color{blue}\onlinecite{Totsuka_97,Chitra_Giamarchi_97,Giamarchi_etal_18b}}].

The very powerful and accurate analytical tool to study spin chains and, in particular, the spin-$1/2$ dimerized Heisenberg chain
is the continuum-limit bosonization approach. A method is well known and discussed in detail in many excellent
reviews and books. Therefore, below we briefly sketch the most relevant steps and bosonization conventions, while for
technical details we refer the reader to the corresponding references [{\color{blue}\onlinecite{GNT_book_98, Giamarchi_book_04, Cabra_Pujol_04}}]. 
To obtain the continuum version of the Hamiltonian ({\color{blue}\ref{XXZ_w_Modulated_DMI_Rotated}}), we use the standard bosonization expression of the spin operators [{\color{blue}\onlinecite{GNT_book_98}}]
\begin{eqnarray}
\tau_{n}^{z} &\simeq&  \sqrt{\frac{K}{\pi}} \partial_x \phi (x)+
(-1)^n
 \frac{{\it a}}{\pi \alpha} \sin\sqrt{4\pi K}\phi (x) \, ,\label{Tau-z_bos}\\
 \tau^{\pm}_{n} &\simeq& \frac{{\it b}}{\pi\alpha}
\cos(\sqrt{4\pi K}\phi)\,e^{\pm i \sqrt{\pi/ K}
\theta}\nonumber\\
& -&(-1)^{n}\frac{{\it c}}{\pi\alpha}\,e^{\pm i\sqrt{\pi/K}\theta}\,
.\label{Tau-pm_bos}
\end{eqnarray}
Here $\phi(x)$ and $\theta(x)$ are dual bosonic fields, $\partial_t
\phi =  u\partial_x \theta $, and satisfy the following
commutational relation
\begin{eqnarray}
\label{regcom}
&& [\phi(x),\theta(y)]  = i\Theta (y-x)\,,  \nonumber\\
&& [\phi(x),\theta(x)]  =i/2\, .
\end{eqnarray}
Here the non-universal real constants {\it a}, {\it b} and {\it c}
depend smoothly on the parameter $\gamma^{\ast}=J_{z}/J^{\ast}$, are of the order
of unity at $\gamma^{\ast}=0$ [{\color{blue}\onlinecite{Hikihara_Furusaki_98}}] and are
expected to be nonzero everywhere at $|\gamma^{\ast}| < 1$. The
Luttinger liquid parameter is known within the critical line $ -1 <
\gamma^{\ast} < 1$ to be [{\color{blue}\onlinecite{Luther_Peschel_75}}]
\bea K &=& \frac{\pi}{2\arccos\left(-\gamma^{\ast}\right)}\, .
\label{K} \eea
Thus the parameter $K$ decreases monotonically from its maximal
value $K \to \infty$ at $\gamma^{\ast} \to -1$ (ferromagnetic
instability point), is equal to unity at $\gamma^{\ast} =0$
($J_{z}=0$) and reaches the value $K=1/2$ at $\gamma^{\ast} =1$
(isotropic antiferromagnetic chain). In the case of dominating Ising
type anisotropy, at $\gamma^{\ast} >1$, $K<1/2$.

Using ({\color{blue}\ref{Tau-z_bos}})-({\color{blue}\ref{Tau-pm_bos}}) we finally obtain for
 the initial lattice Hamiltonian ({\color{blue}\ref{XXZ_w_Modulated_DMI_Rotated}}):
\begin{eqnarray}  \label{H_XXZ+D1_bos_3SG}
{\cal H}&=&\int dx\,\Big[\frac{u}{2}(\partial_{x}\phi)^2+\frac{u}{2}(\partial_{x}\theta)^2 -H \sqrt{\frac{K}{\pi}}\partial_{x}\phi\nonumber\\
&&-\frac{\delta J^{*}}{\pi\alpha} \cos\sqrt{4\pi K}\phi - \frac{H_{1}}{\pi\alpha} \sin\sqrt{4\pi K}\phi\nonumber\\
&& +\frac{J_{z}}{2\pi^{2}\alpha^{2}}\cos\sqrt{16\pi K}\phi \,\Big]\, ,
\end{eqnarray}
where $u \simeq J^{\ast}/K$ is the velocity of spin excitation. It is instructive to unify the first two nonlinear terms with equal arguments into one and rewrite the Hamiltonian in the standard form of the double-frequency sine-Gordon model
\begin{eqnarray}  \label{H_XXZ+D1_bos_DSG}
&&\hspace{0mm}{\cal H}=\int dx \Big[\frac{u}{2}(\partial_{x}\phi)^2+\frac{u}{2}(\partial_{x}\theta)^2 
-H \,\frac{\beta}{2\pi}\partial_{x}\phi \nonumber\\
&&\hspace{+5mm} -\frac{\Delta_{0}}{\pi\alpha^{2}} \cos\beta(\phi-\phi_{0}) + \frac{M_{0}}{\pi\alpha^{2}}\cos2\beta\phi \,\Big]\, . 
\end{eqnarray}
where $\beta=\sqrt{4\pi K}$, $M_{0}=J_{z}/2\pi$, 
\begin{eqnarray}  \label{Delta_0_Bos}
\Delta_{0}&=&\sqrt{H_{1}^{2}+(D_{0}D_{1}/J^{\ast})^{2}},
\end{eqnarray}
and 
\begin{eqnarray}
\phi_{0} = \arctan\left(H_{1}/\delta J^{\ast}\right)\, .
\end{eqnarray}

The scaling dimensions of the first two {\em cosine} terms $d=dim[\cos(\beta\phi)]=\beta^{2}/4\pi=K$, while the scaling dimension of the third {\em cosine} term is $d^{\ast}=dim[\cos(2\beta\phi)]=4K$. Each of two {\em cosine} terms in ({\color{blue}\ref{H_XXZ+D1_bos_DSG}}) become relevant in the parameter
range where the corresponding scaling dimensionality $d \leq 2$ or
$d^{\ast}\leq 2$. Using ({\color{blue}\ref{K}}) we find that $d \leq 2$, i.e.\ the
first {\em cosine} term in ({\color{blue}\ref{H_XXZ+D1_bos_DSG}}) is relevant, at
\begin{eqnarray} \label{gamma_c1} 
\gamma^{\ast}>\gamma^{\ast}_{c_1}=-\sqrt{2}/2\, ,
\end{eqnarray}
while  the second {\em
cosine} term in ({\color{blue}\ref{H_XXZ+D1_bos_DSG}}), for $\gamma^{\ast} > 1$.
Therefore, at $H=0$ the antiferromagnetic sector of the phase diagram at 
$\gamma^{\ast} > -1$ can be divided in two segments of the model parameter range: 
the gapless LL sector at $-1 <\gamma^{\ast} \leq \gamma^{\ast}_{c_1}$
and the gapful sector at $\gamma^{\ast} > \gamma^{\ast}_{c_1}$. 

\subsection{The LL sector $-1<\gamma^{\ast}< \gamma^{\ast}_{c_1}$}
\label{subsec:B}

At  $-1<\gamma^{\ast}< \gamma^{\ast}_{c_1}$, both cosine terms in ({\color{blue}\ref{H_XXZ+D1_bos_DSG}}) are
irrelevant and can be neglected. The gapless long-wavelength
excitations of the anisotropic spin chain are described by the
standard Gaussian theory with the Hamiltonian
\begin{eqnarray}\label{SpinChainBosHam_Gauss}
{\cal H}_{0}& = &u\int dx \, \Big[\,\frac{1}{2}(\partial_x \phi)^{2}
 + \frac{1}{2}(\partial_x
\theta)^{2}\,\Big].
\end{eqnarray}
In this {\em critical LL phase}, all correlations show a power-law decay, with indices, smoothly depending on the parameter $K$ [{\color{blue}\onlinecite{GNT_book_98}}]. Because the LL parameter $K$ depends only on the anisotropy parameter $\gamma^{\ast}$, the very presence of magnetic field does not change the values of the parameter $\gamma^{\ast}$ where the {\em conine} term becomes relevant, however, as we show below, magnetic field substantially influence properties of the model in the gapped phase and at leads to the extension of the LL sector.

\subsection{The gapped sector $\gamma^{\ast}_{c_1}< \gamma^{\ast} \leq 1$}
\label{subsec:C1}

At $H=0$ and $\gamma^{\ast}_{c_1} < \gamma^{\ast} \leq 1$ the first cosine in ({\color{blue}\ref{H_XXZ+D1_bos_DSG}}) is 
relevant perturbation, while the double-frequency cosine term remains 
irrelevant and can be neglected. In this case infrared properties of
the system are described by the standard sine-Gordon (SG) model
\begin{eqnarray}  \label{H_XXZ+D1_bos_SG}
{\cal H}&=&\int dx
\,\Big[\frac{u}{2}(\partial_{x}\phi)^2+\frac{u}{2}(\partial_{x}\theta)^2\nonumber\\
&-&\frac{\Delta_{0}}{\pi\alpha^{2}} \cos\beta(\phi-\phi_{0})  \Big].
\end{eqnarray}
At  $\gamma^{\ast}=\gamma^{\ast}_{c_1} \simeq -0.7$ the Berezinskii-Kosterlitz-Thouless (BKT)
[{\color{blue}\onlinecite{BKT_tansition}}] quantum phase transition takes place in the
ground state of the system, the excitation gap opens and remains finite in the whole
region $\gamma^{\ast}>\gamma^{\ast}_{c_1}$. From the exact solution of the quantum sine-Gordon model
[{\color{blue}\onlinecite{SG_exact_Sol,Zamolodchikov_95}}] it is known that for
arbitrary finite $\Delta_{0}$ the gapped excitation spectrum of the
Hamiltonian Eq.\ ({\color{blue}\ref{H_XXZ+D1_bos_SG}}) at $-0.7 <\gamma^{\ast}\leq 0$ consists of solitons and antisolitons with
masses
\begin{equation}  \label{Soliton_Mass}
{\cal M}_{sol} \sim
\left(\Delta_{0}/J^{\ast}\right)^{1/(2-K)},
\end{equation}
while at $0 < \gamma^{\ast}\leq 1$  in addition, 
also of soliton-antisoliton bound states ("breathers") with the
lowest breather mass
\begin{equation}  \label{Breathers_Mass}
{\cal M}_{br}=2{\cal M}_{sol}\sin\left(\frac{\pi K}{4-2K}\right)\, .
\end{equation}
Thus, in the whole parameter range $0 < \gamma^{\ast}\leq 1$ the
soliton mass ${\cal M}_{sol}$ is the energy scale that determines
the size of the spin excitation gap.

At the BKT phase transition point, the excitation gap is exponentially small  
\begin{equation}  \label{Excitation_Gap_1}
\Delta_{exc}\sim J^{\ast}
\exp\left(-1/(\gamma^{\ast}-\gamma^{\ast}_{c_1})\right)\, ,
\end{equation}
it smoothly increases with increasing $\gamma^{\ast}$, and at $\gamma^{\ast}=0$
\begin{equation}  \label{Excitation_Gap_1}
\Delta_{exc}=2J^{\ast}{\cal M}_{sol}=2\Delta_{0}\, ,
\end{equation}
in a perfect agreement with results obtained in the Sec.~\ref{sec:TheModel} (see Eq.({\color{blue}\ref{Gap_XX_D1}})). 

Generation of a gap in the excitation spectrum leads to suppression of
fluctuations in the system and the  $\phi$ field is condensed in one
of its vacua corresponding to the minimum of the dominating potential energy term 
$V(\phi)=-\Delta_{0}\cos\sqrt{4\pi K}(\phi-\phi_{0})$ [{\color{blue}\onlinecite{Mutalib_Emery_86}}]
\begin{eqnarray}  \label{Ordered-Field_1}
\langle 0|\,\phi\,|0 \rangle  &=& \phi_{0}+2n\pi,  \hskip 1.0cm \textrm{ at \quad $\Delta_{0} > 0$}\, .
\end{eqnarray}
Using the vacuum expectation value of the $\phi$ field ({\color{blue}\ref{Ordered-Field_1}}) one easily obtains that 
in the ground state, the system is characterized by the LRO pattern of the on-site  
staggered magnetization with amplitude
\begin{eqnarray}  \label{Order_parameters_tau_m}
&& M_{st}=(-1)^{n} \langle 0|\tau^{z}_{n}|0\rangle \sim \sin\sqrt{4\pi
K}\phi_{0}\, .
\end{eqnarray}
Moreover, if we consider the link-located degrees of freedom, using
({\color{blue}\ref{Tau-z_bos}})-({\color{blue}\ref{Tau-pm_bos}}) one obtains that in the ground state of the system
the staggered component of the $\tau$-spin chirality operator given by
\begin{eqnarray}
&&\kappa_{n}^{(\tau)}=-i(-1)^{n}\langle 0|\left(\tau^{+}_{n}\tau^{-}_{n+1}-h.c.\right)|0\rangle\rightarrow\nonumber\\
&&\simeq \frac{{\it b}}{\pi\alpha}\sin(\sqrt{4\pi K}\phi_{0})\,
\label{mapping_Kappa-Tau}
\end{eqnarray}
together with staggered parts of the $\tau$-spin longitudinal and transverse nearest-neighbour spin-exchange operators
\begin{eqnarray}
&&{\cal D}_{\perp}^{(\tau)}(n)=\frac{(-1)^{n}}{2}\langle 0|\left(\tau^{+}_{n}\tau^{-}_{n+1}+h.c.\right)|0\rangle 
\nonumber\\
&& \hspace{+15mm}\sim 
\frac{{\it a}}{\pi\alpha}\cos(\sqrt{4\pi K}\phi_{0})\label{mapping_Epsilon_perp_Tau}\\
&&{\cal D}_{z}^{(\tau)}(n)=(-1)^{n}\langle 0|\tau^{z}_{n}\tau^{z}_{n+1}| 0 \rangle \nonumber\\
&& \hspace{+15mm}\sim \frac{b}{\pi\alpha}\cos(\sqrt{4\pi K}\phi_{0})\label{mapping_Epsilon_paralel_Tau}
\end{eqnarray}
show the long-range ordered patterns in the ground state of the effective $\tau$-spin model.

Using ({\color{blue}\ref{S2Tau}}), from
({\color{blue}\ref{Order_parameters_tau_m}})-({\color{blue}\ref{mapping_Epsilon_paralel_Tau}}) we obtain, that in the
gapped phase the initial spin chain shows a long-range dimerization
order
\begin{eqnarray}
&& \hspace{-3mm}\frac{1}{N}\sum_{n}(-1)^{n}
\la\,\textbf{S}_{n}\cdot\textbf{S}_{n+1} \, \ra \sim
(\cos\vartheta_{+}-\cos\vartheta_{-})\,\epsilon \,,
\label{S-Dimerization-Pattern}
\end{eqnarray}
which coexists with the LRO  pattern of the alternating
spin chirality vector
\begin{eqnarray}
&& \hspace{-3mm}\frac{1}{N}\sum_{n}(-1)^{n}\la \,\kappa^{z}_{n}\ra
\sim
(\sin\vartheta_{+}-\sin\vartheta_{-})\, \, \kappa^{z}_{0}.
\label{S-Chirality-Pattern}
\end{eqnarray}
and of the staggered magnetization
\begin{eqnarray}  \label{Order_parameters_tau_S}
&&\frac{1}{N}\sum_{n}(-1)^{n} \langle S^{z}_{n} \rangle =\, \,m .
\end{eqnarray}

\subsection{The gapped sector $ \gamma^{\ast} > 1$}

At $\gamma^{\ast} > 1$ the second {\em cosine} term in ({\color{blue}\ref{H_XXZ+D1_bos_DSG}}) becomes relevant 
and the effective continuum-limit version of the initial lattice spin model ({\color{blue}\ref{XXZ_w_Modulated_DMI_Rotated}}) is given by the double-frequency sine-Gordon (DSG) model [{\color{blue}\onlinecite{DSG_Book_80}}]
\begin{eqnarray}  \label{H_XXZ+D1_bos_DSG_2}
&&\hspace{-1mm}{\cal H}=\int dx
\,\Big[\frac{u}{2}(\partial_{x}\phi)^2+\frac{u}{2}(\partial_{x}\theta)^2 
\nonumber\\
&&\hspace{-5mm}-\frac{\Delta_{0}}{\pi\alpha^{2}} \cos\beta(\phi-\phi_{0}) + \frac{M_{0}}{\pi\alpha^{2}}\cos(2\beta\phi) \,\Big]\, .  
\end{eqnarray}
The DSG model describes an interplay between two relevant perturbations to the Gaussian conformal field theory with the ratio of their scaling dimensions equal to 4. Infrared properties of the DSG model are determined by an interplay between these two relevant sources of perturbations to the Gaussian conformal field theory. Acting separately, each leads to the pinning of the field $\phi$ in corresponding minima, and depending on sign and amplitudes of the model parameters, $\Delta_{0}$ and $J_{z}$, these minima either exclude each other or support synergic ordering [{\color{blue}\onlinecite{DSG_1,DSG_2a}}]. 

In absence of the staggered component of the magnetic field $\phi_{0}=0$ the spin dimerization, provided by the alternating DMI interaction and the staggered magnetization, supported by the Ising part of the exchange interaction have different parity symmetries, the field configurations which minimize one perturbation do not minimize the other. This destructive competition between possible sets of vacuum configurations of the two cosine terms is resolved via the Ising type QPT in the ground state at $J^{\ast}_{z} \sim 1+ D_{0}D_{1}/J^{\ast}$ from a dimerized phase at $J_{z}<J^{\ast}_{z}$ into the composite phase with coexisting {\em dimer} and {\em antiferromagnetic} order at $J_{z}>J^{\ast}_{z}$ [{\color{blue}\onlinecite{Japaridze_etal_19a}}].   The very presence of the corresponding QPT can already be traced performing minimization of the potential
\begin{eqnarray}\label{DSG_Potential}
 V(\phi)&=&-\Delta_{0}\cos\beta\phi + M_{0}\cos2\beta\phi\, .
\end{eqnarray}
The Ising transition corresponds to the crossover from a double well
to a single well profile of the potential ({\color{blue}\ref{DSG_Potential}}).
indeed, one can easily obtain, that at $4M_{0}<\Delta_{0}$ the vacuum
expectation value of $\phi$ field which minimizes $V(\phi)$
is $\langle\phi\rangle=0$ and therefore in this case only the
dimerized phase is realized in the ground state. However, at
$4M_{0}>\Delta_{0}$ the $\phi$ field is condensed in the minima
\begin{eqnarray}  \label{Ordered-Field_2}
\beta\langle \phi \,\rangle = 
\pm\,\arccos\left(\Delta_{0}/4M_{0}\right)+2\pi n \,
\end{eqnarray}
and, as the result, the ground state of the $\tau$-spin system, in addition to the dimerization pattern the long-range antiferromagnetic order is present.
\begin{figure}[t]
\centerline{\psfig{file=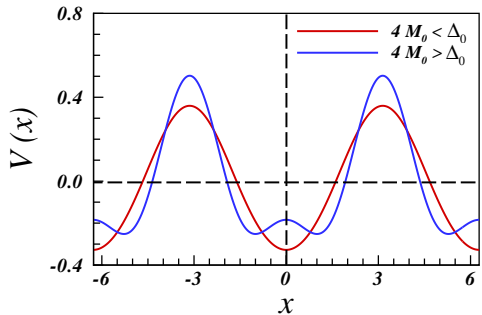,width=2.7in}}
\caption{The DSG potential $V(x)=-\Delta_{0}\cos x + M_{0}\cos2x$  for $4M_{0}<\Delta_{0}$ (red) and for $4M_{0}>\Delta_{0}$
(blue). We have taken a DMI as $D_ 0 = 0.7$ and $D _1 = 0.3$ for $\gamma=0.1$ (red) and $\gamma=1.0$ (blue). It is evident that at $4M_{0}>\Delta_{0}$ the structure of minima is changed. Appear new transitions between new set of minima corresponding to solitons capturing topological charge (spin) equal to $\phi_{0}/2\pi$ and $1-\phi_{0}/2\pi$.  }
\label{Vacua_System_2}
\end{figure}

The Ising transition at $J=(J_{z})_{c_2}$ is also displayed in the character of excitations. In absence, as well as for the weak double-frequency cosine term ($4M_{0}<\Delta_{0}$) the vacuum expectation value of the field is determined by the set of minima of the $-\Delta_{0}\cos\beta\phi$ term given by ({\color{blue}\ref{Ordered-Field_1}}). Therefore the spin of the  elementary excitation i.e the quantum number captured by the soliton kink between nearest minima
\begin{eqnarray}  \label{Soliton_QN1}
S=\frac{\beta}{2\pi}\int_{-\infty}^{+\infty}dx\partial_{x}\phi(x)=1,
\end{eqnarray}
and corresponds to the magnon type excitation.

In the case of doubled potential there appear solitons with different masses (see Fig. {\color{blue}\ref{Vacua_System_2}}), one corresponds to the kink between two split minima, and the second, of the larger mass, corresponds to the kink over the high potential. The first soliton captures the spin
\begin{eqnarray}  \label{Soliton_QN1}
s_{1}=\frac{\beta}{2\pi}\int_{-\infty}^{+\infty}dx\,\partial_{x}\phi(x)=
\frac{1}{\pi}\arccos\left(\Delta_{0}/4M_{0}\right),
\end{eqnarray}
while the second, 
\begin{eqnarray}  \label{Soliton_QN1}
s_{2}=\frac{\beta}{2\pi}\int_{-\infty}^{+\infty}dx\partial_{x}\phi(x)=1-s_{1}\, . 
\end{eqnarray}
In absence of the dimerization ($\delta=0$),  masses and spins of both soliton $s_{1}=s_{2}=1/2$, which corresponds to the spinon type excitations in the Heisenberg chain in the gapful sector. 
 
\begin{figure}[t]
\centerline{\psfig{file=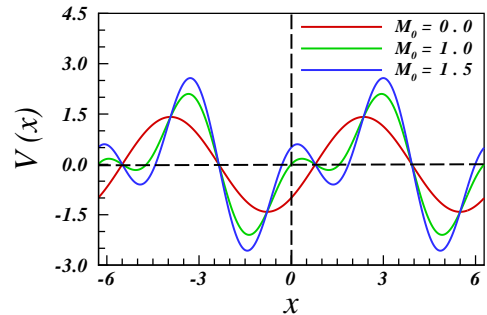,width=2.7in}}
\caption{(color online) The DSG potential $V(x)=-\Delta_{0}\cos x +H_{1}\sin x + M_{0}\cos2x$ for values of the parameters $\Delta_{0}=M_{0}=1$ and for different values of the parameter $M_{0}=0$ (red), $M_{0}=1$ (green) and $M_{0}=1.5$ (blue). It is clear that the distance between absolute minima of the potential remain unchanged. }
\label{Vacua_System_3}
\end{figure}

In the presence of finite staggered magnetic field  the $\phi \rightarrow -\phi$ symmetry of the model is broken for arbitrary 
$H_{1}\neq 0$. Already in the absence of the double frequency cosine term, the minimum of the potential is reached at $\phi=\phi_{0}+2l\pi$. Presence, at  $\gamma^{\ast}>1$, of the second {\em relevant cosine} term in the Hamiltonian results to shift off the minima positions, make minima deeper, but {\em does not change the structure of the vacua} (see Fig. {\color{blue}\ref{Vacua_System_3}} ), the distance between minima remain unchanged and equal to $2\pi$ and therefore the ground state and character of excitations is not influenced. 

Thus, in constant with the $H_{1}\neq 0$ case {\color{blue}\cite{Japaridze_etal_19a}}, at the $\gamma^{\ast}>1$ point the Ising transition is absent and in the whole parameter range $\gamma^{\ast}>\gamma^{\ast}_{c_1}$ the ground state is characterized by the composite order with coexisting dimer, alternating chirality and antiferromagnetic order, massive excitations have the similar nature, are magnons and capture spin $S=1$ and the excitation gap smoothly increases in the whole area  $\gamma^{\ast}>\gamma^{\ast}_{c_1}$ with weak anomaly at the {\em crossover} point at $\gamma^{\ast}>1$.

To summarize this section, the staggered magnetic field 
\begin{itemize}
\item{  leads to the presence of the composite ordered phase in the whole parameter range 
$\gamma^{\ast} >\gamma^{\ast}_{c_1}$;}
\item{ increases the bare value of the soliton mass $\Delta_{0}$ i.e leads to enlargement  of the excitation gap 
in the whole parameter range 
$\gamma^{\ast} >\gamma^{\ast}_{c_1}$} .
\end{itemize}

\begin{figure*}[t]
\centerline{\psfig{file=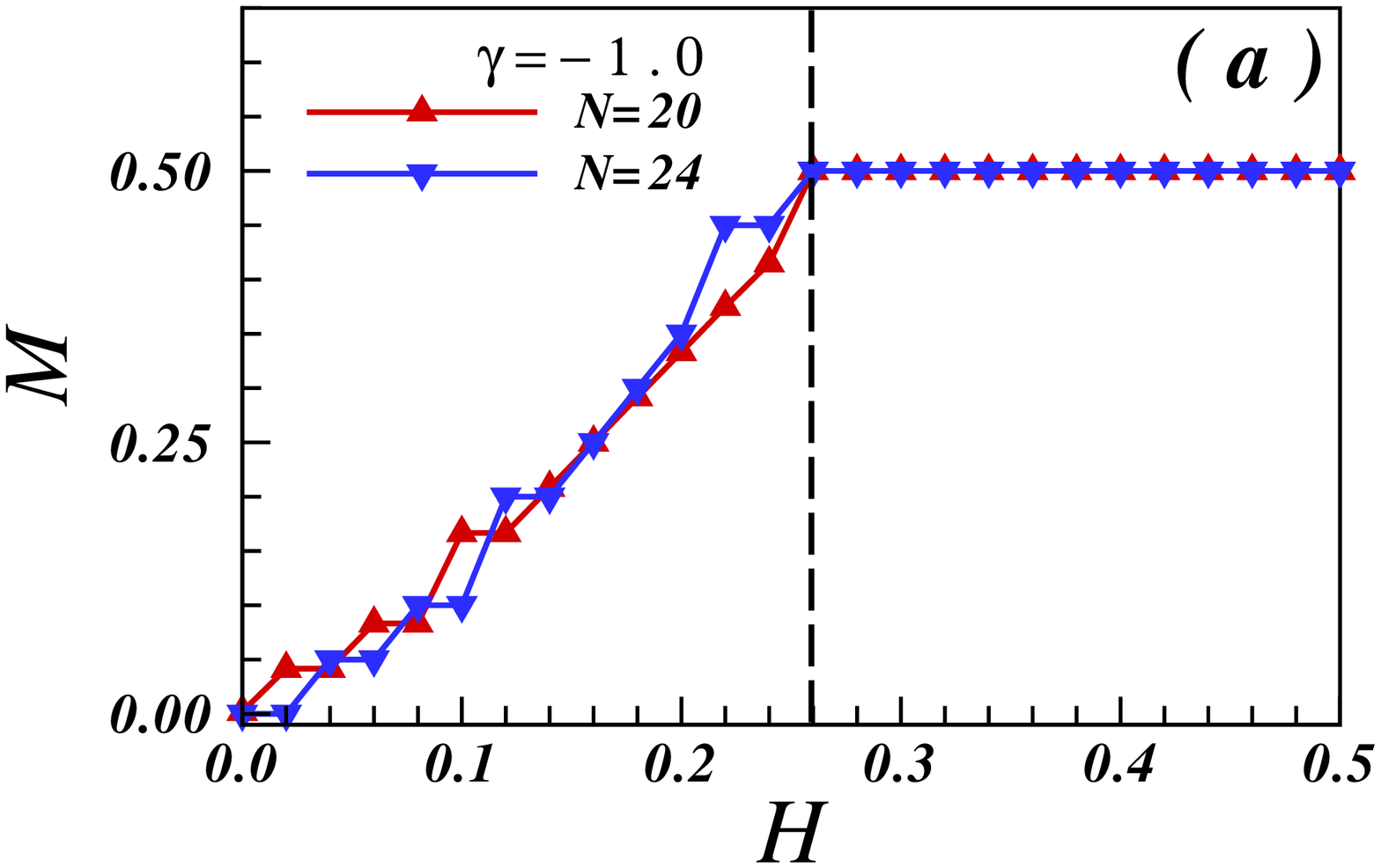,width=2in} \psfig{file=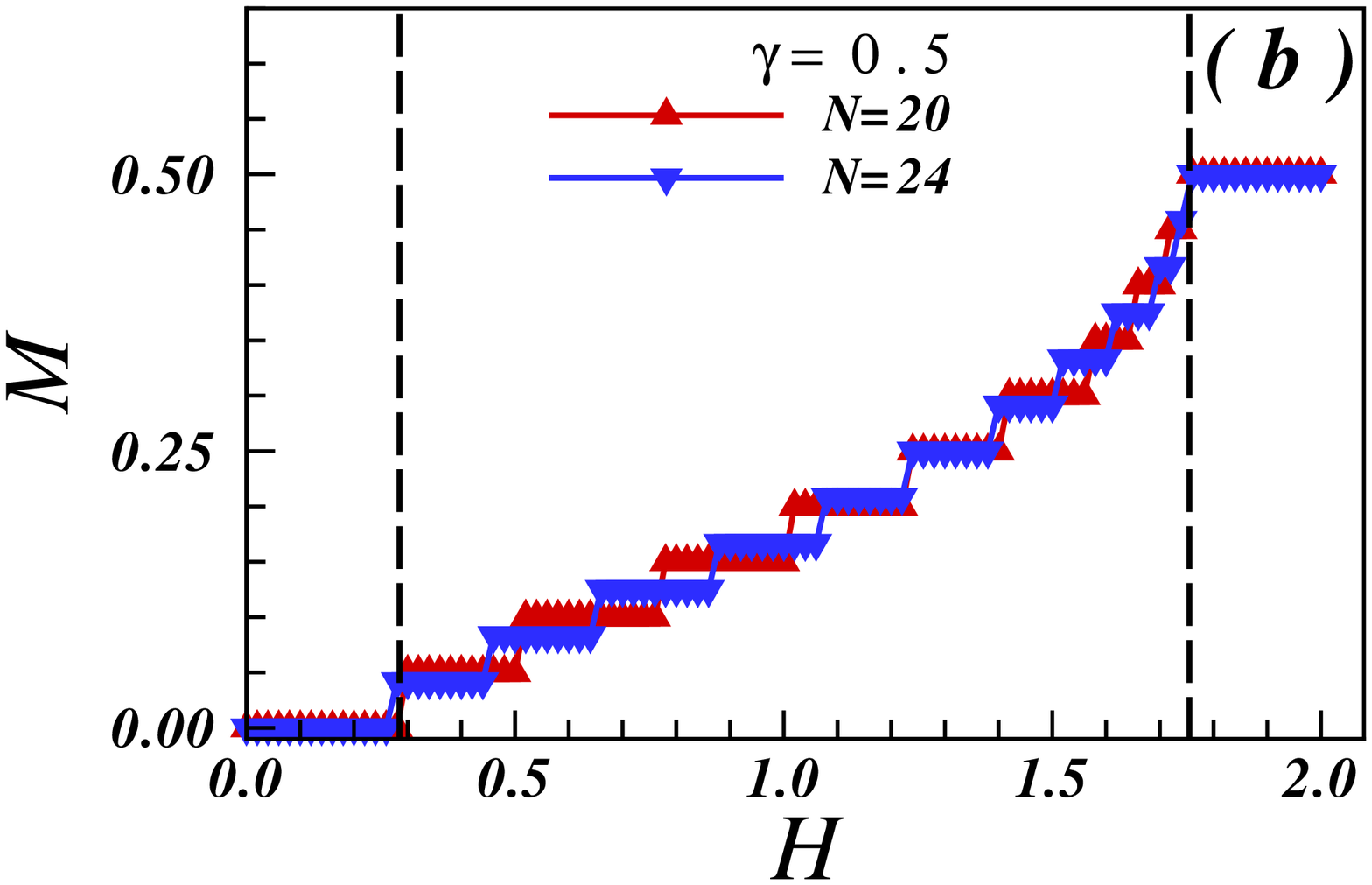,width=2in}\psfig{file=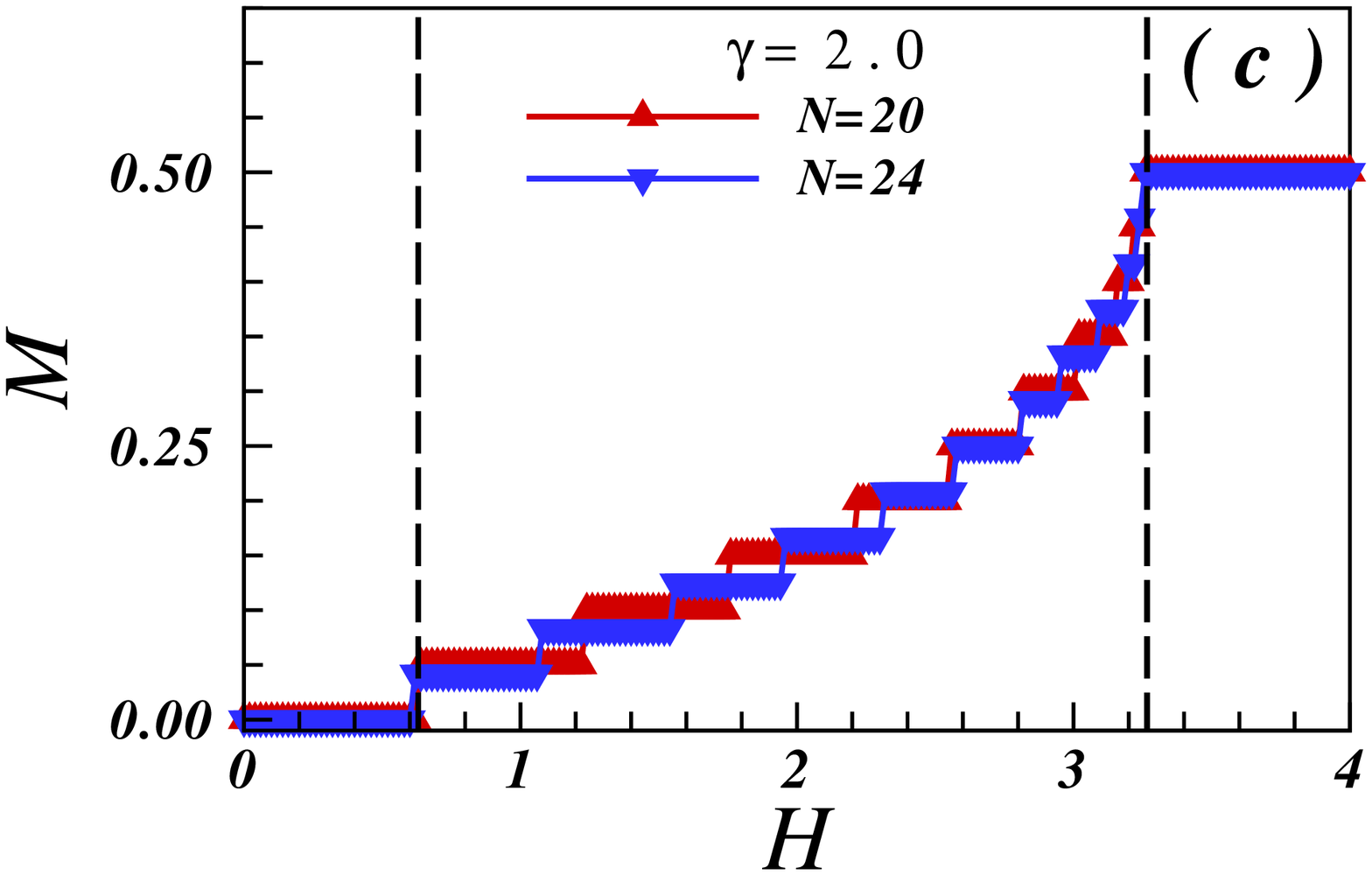,width=2in}}
\caption{(color online) Magnetization versus the uniform magnetic field for sizes $N=20,24$ provided for a DMI as $D_ 0 = 0.7$ and $D _1 = 0.3$ for  (a) $\gamma=-1.0$, (b)  $\gamma=0.5$, and (c) $\gamma=2.0$.  The vertical black dashed lines hint the critical uniform fields. }
\label{Mag-Lanc}
\end{figure*}

\section{The C-IC transition}
At $H \neq 0$ the Hamiltonian ({\color{blue}\ref{H_XXZ+D1_bos_DSG}}) is easily recognized as
the  Hamiltonian for the commensurate-incommensurate (C-IC) phase transition [{\color{blue}\onlinecite{C_IC_transition_1,C_IC_transition_2}}] which has been intensively studied
in the past using bosonization approach [{\color{blue}\onlinecite{JN_79}}] and the Bethe
ansatz technique [{\color{blue}\onlinecite{JNW_84}}]. Below we use the results obtained in these studies to 
give brief sketch of the phase diagram.

At $ H \neq 0$, the presence of the gradient term in the Hamiltonian ({\color{blue}\ref{H_XXZ+D1_bos_DSG}}) makes it necessary to consider the ground state of the sine-Gordon model in sectors with nonzero topological charge. The effective chemical potential
\begin{eqnarray}  \label{Grad term}
\sim -H \sqrt{\frac{\beta}{4\pi^{2}}}\partial_{x}\phi\, , 
\end{eqnarray}
tends to change the number of particles in the ground state i.e. to create a finite and uniform density of solitons. It is clear that the
gradient term in (\ref{H_XXZ+D1_bos_DSG}) can be eliminated by a gauge transformation
\begin{eqnarray}  \label{Shift}
\phi \rightarrow \phi +\frac{H\beta}{2\pi}\,x \, , 
\end{eqnarray}
however, this immediately implies that the vacuum distribution of the field $\phi$ will be shifted with respect to the corresponding minima. Competition between the uniform magnetic field, which supports the ground state characterized by the finite value of the gradient of the $\phi(x)$ field and of the nonlinear {\em cosine} terms, which prefer pinned in one of the potential minima 
the constant value of the $\phi(x)$ field,  is resolved as a continuous C-IC phase transition from a gapped state at $H < H_{c}=M$ to a gapless (paramagnetic) phase at $H>H_{c}$, where $M$ is the  mass of the soliton of the corresponding SG or DSG theory [{\color{blue}\onlinecite{C_IC_transition_1}}]. 

As usual in the case of C-IC transition, in the close proximity of the transition point, 
$H > H_{c}$ $(H-H_{c}) \ll H_{c}$, the magnetization shows a square-root behavior 
\begin{eqnarray}  \label{Magnetization}
M(H) \sim (H-H_{c})^{1/2} \
\end{eqnarray}
and the magnetic susceptibility a square-root divergence
\begin{eqnarray}  \label{Susceptibility}
\chi(H) \sim (H-H_{c})^{-1/2}\, .
\end{eqnarray}

At $H > H_{c}$ the excitation spectrum becomes gapless, all correlations show a power-law decay with exponents dependent on 
$\gamma^{\ast}$ and $H$ and the system exhibits properties of the magnetized spin LL. 

Thus in the presence of alternating magnetic field the ground-state phase diagram depending on the model parameters consist of the following two sectors:

\begin{itemize}
\item{  The gapless LL phase with finite magnetization at $H>H_{c}$;} 
\item{ The gapped phase with composite order, characterized by the coexistence of  the LRO dimerization, alternating spin chirality (spin current), and antiferromagnetic patterns  $H<H_{c}$.} 
\item{ The value of the critical field $H_{c}=M$ is determined by the mass of the soliton of the corresponding SG or DSG theory.}
\item{ The soliton mass is determined by its bare value $\Delta_{0}$ and the effective anisotropy parameter $\gamma^{\ast}$.} 
\item{ The effective anisotropy parameter $\gamma^{\ast}$ depends on both symmetric ($J$) and asymmetric ($D_{0},D_{1}$.) 
components of spin exchange;}
\item{ The bare value of the soliton mass  $\Delta_{0}$ is additively contributed by the staggered component of the magnetic field and alternating DMI.}

\end{itemize}


\section{Numerical results for an arbitrary value of $\gamma$.}

In this section, we present the results of our numerical studies for finite chains. We use the numerical Lanczos technique as one of the most frequently used numerical algorithms in investigating the ground-state phase diagram of low-dimensional spin-1/2 systems. Applying the Lanczos algorithm to the transformed Hamiltonian, we diagonalize numerically finite chains up to $N=28$ spin-1/2 particles and calculate the uniform and the staggered magnetizations for arbitrary values of  $\gamma=J_{z}/J  $.

In Fig.~{\color{blue}\ref{Mag-Lanc}} we have illustrated the magnetization as a function of the uniform magnetic field for a specified value of DMI as $D_0=0.7$ and $D_1=0.3$  in the absence of the staggered magnetic field. As depicted in Fig.~{\color{blue}\ref{Mag-Lanc}(a)}, when the system is put in the gapless LL phase, as soon as the uniform field is exerted the magnetization process starts and the system remains in the LL phase up to a critical uniform field $H_c\simeq 0.26\pm0.01$ where the system goes into the paramagnetic phase. Observed oscillations of the magnetization result from the level crossing between the ground and the excited states of this model in the gapless LL phase.
The magnetization curves started from  the gapped composite $C1$ and $C2$ phases are manifested in Figs.~{\color{blue}\ref{Mag-Lanc}(b)} and {\color{blue}\ref{Mag-Lanc}(c)}. As is observed, the magnetization remains zero up to when the value of the uniform field reaches to the first critical value as $H_{c_1}(\gamma=0.5)\simeq 1.76\pm0.01$ and $ H_{c_1}(\gamma=2.0)\simeq 3.26\pm0.01$,  that are equal to the spin gaps. In complete agreement with our analytical calculations, more increment of the uniform field entails the enhance of the magnetization in such a way that makes it to be saturated in the second critical field as $H_{c_2}(\gamma=0.5)\simeq 0.28\pm0.01$ and $H_{c_2}(\gamma=2.0)\simeq 0.62\pm0.01$.

\begin{figure}[t]
\centerline{\psfig{file=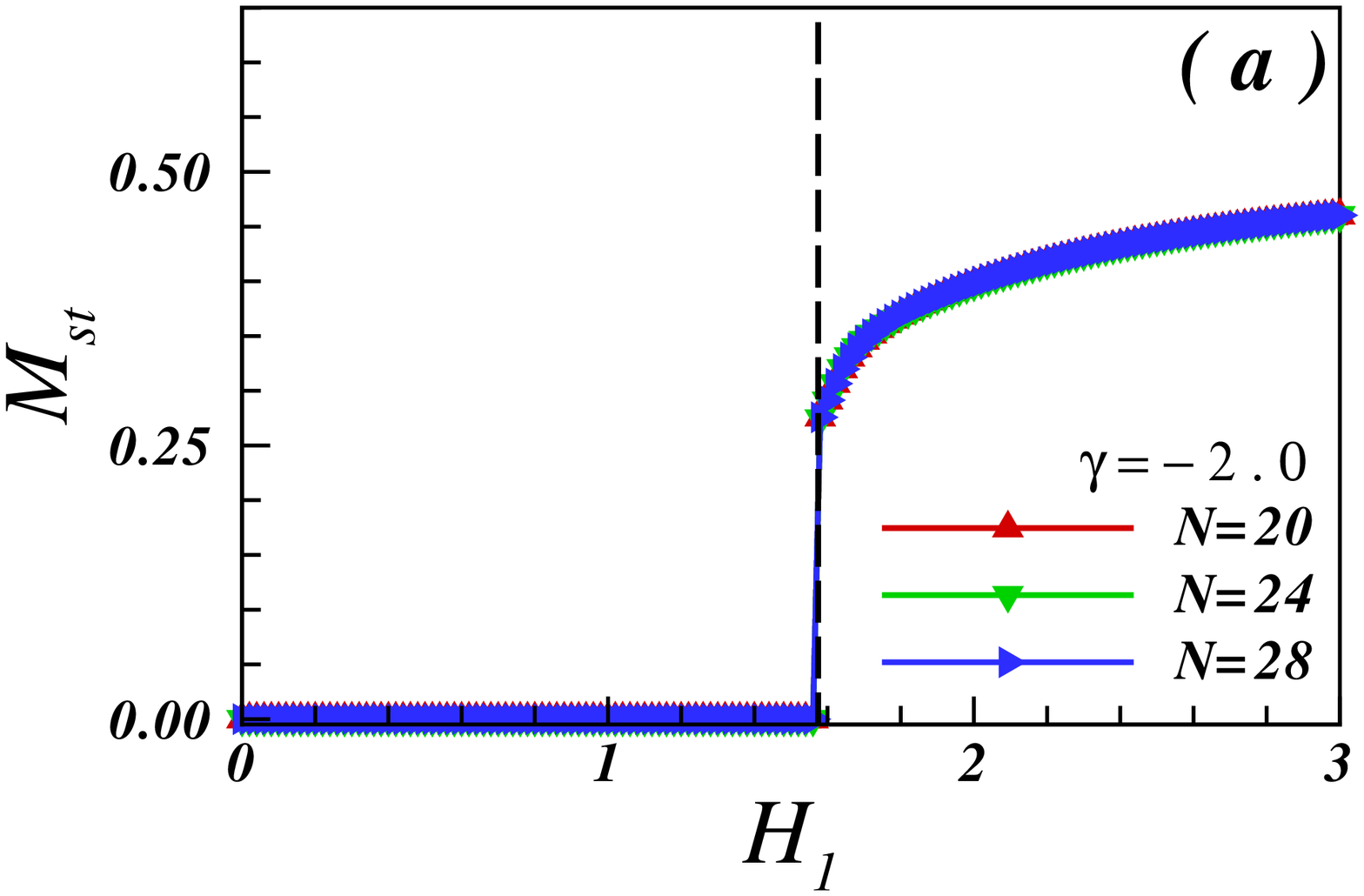,width=1.7in} \psfig{file=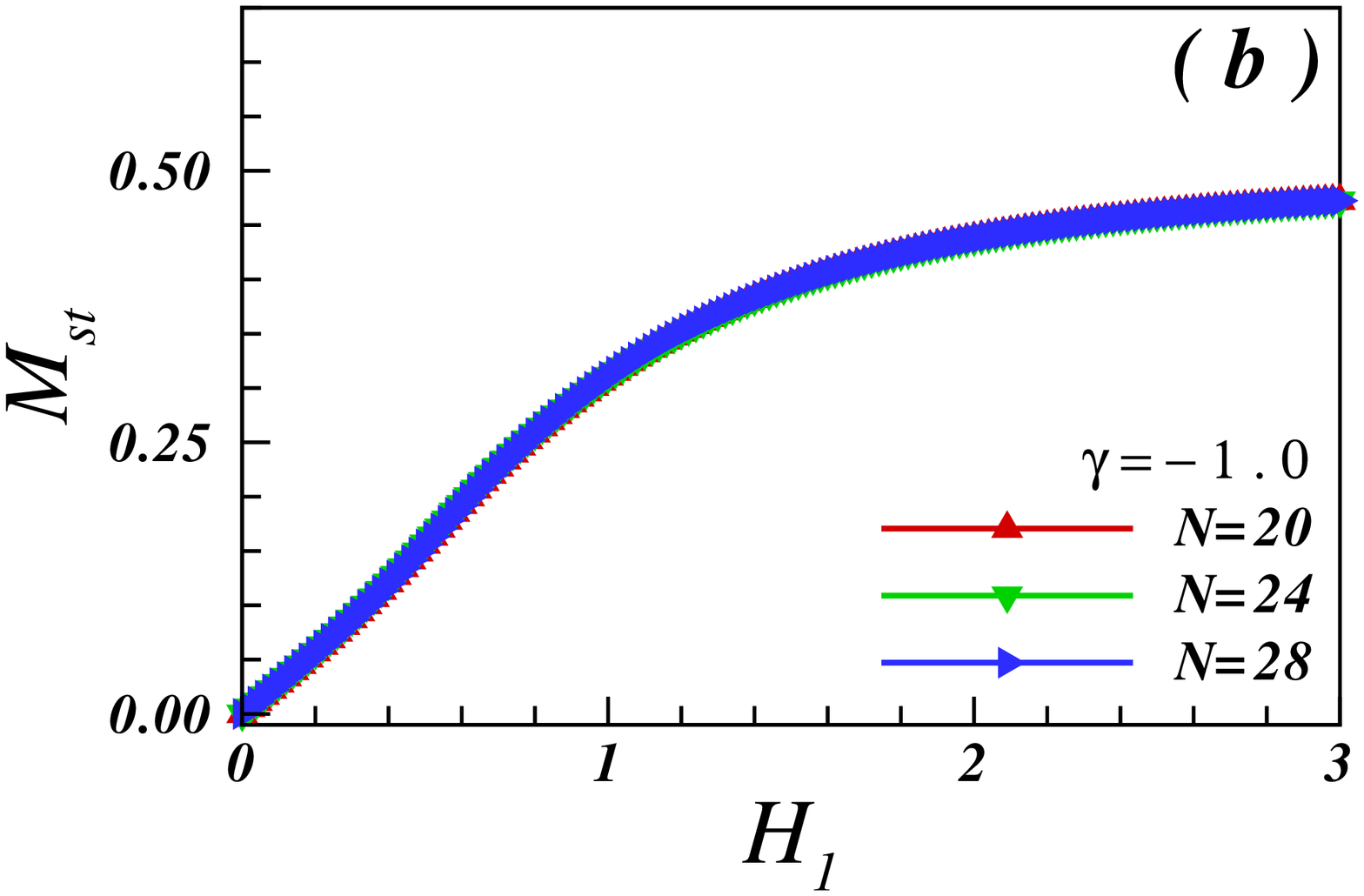,width=1.7in}}
\centerline{\psfig{file=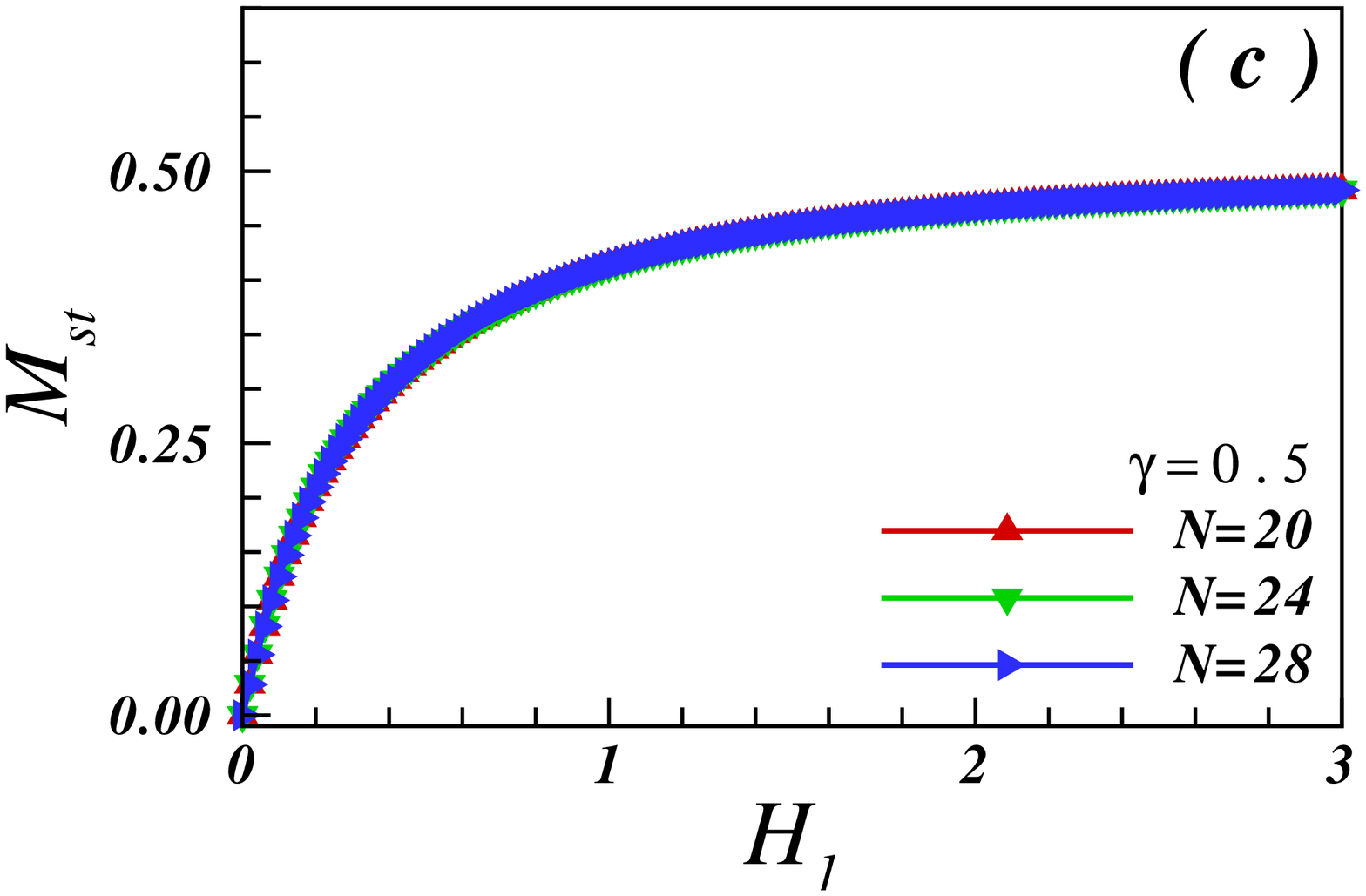,width=1.7in} \psfig{file=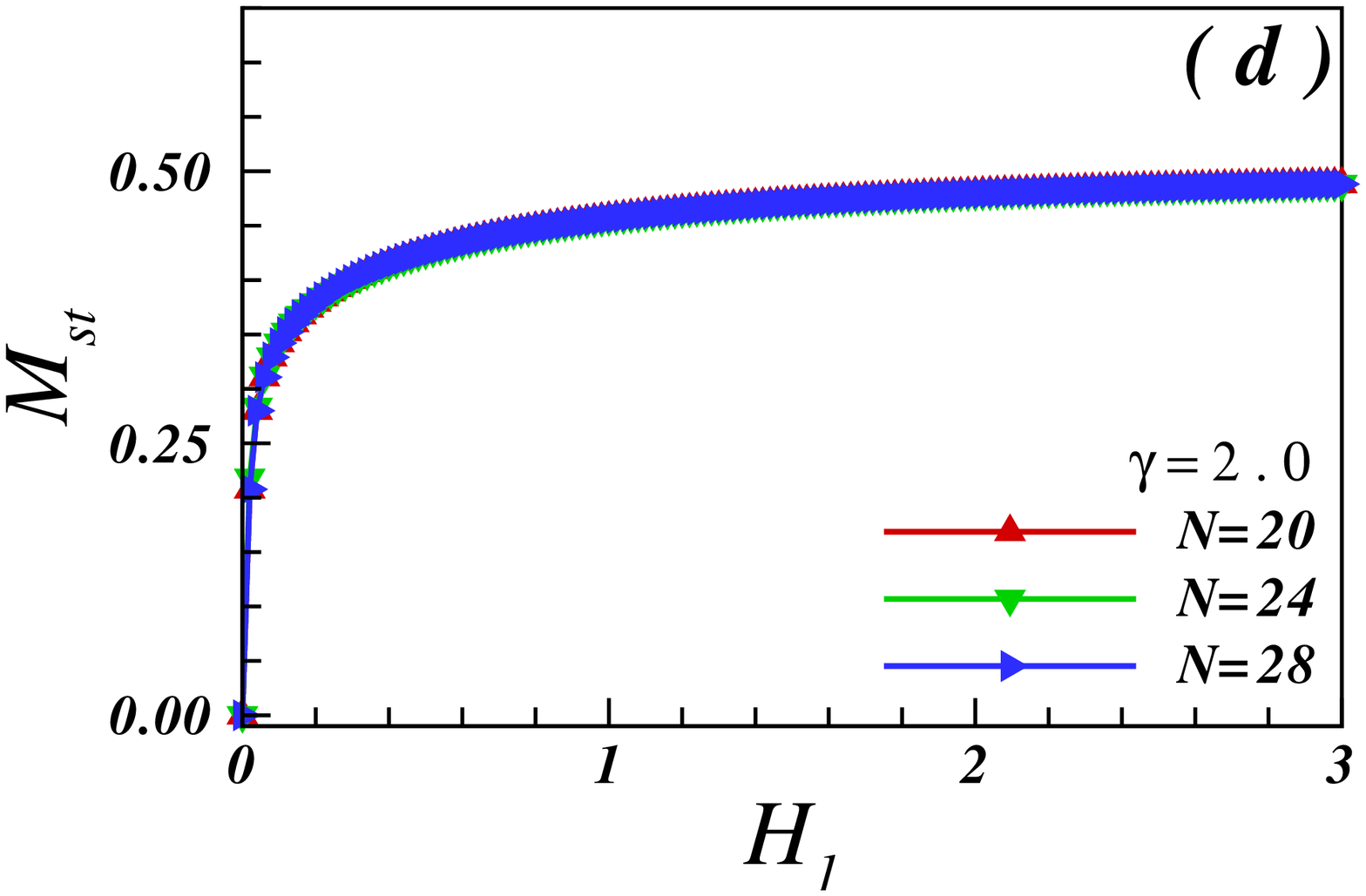,width=1.7in}}
\caption{(color online) Staggered magnetization versus the staggered magnetic field for different sizes $N=20,24,28$  for a DMI as $D_ 0 = 0.7$ and $D _1 = 0.3$ supplied for  (a) $\gamma=-2.0$, (b)  $\gamma=-1.0$,  (c) $\gamma=0.5$, and (d) $\gamma=2.0$. As is viewed in (a), the FM phase resists the transition up to when the staggered field touches a critical value (the vertical black dashed line). }
\label{St-Mag-Lanc}
\end{figure}

We also have studied the effect of the staggered magnetic field on the ground-state phase diagram of the model. Numerical Lanczos results are presented  in Fig.~{\color{blue}\ref{St-Mag-Lanc}} for values of DMI as $D_0=0.7$ and $D_1=0.3$.  Since the staggered field has no effect on the FM state a zero-plateau is observed in the curve of staggered magnetization for $\gamma=-2.0$ [see Fig.~{\color{blue}\ref{St-Mag-Lanc}(a)}].  In the gapless LL region, as is seen in Fig.~{\color{blue}\ref{St-Mag-Lanc}(b)}, the staggered magnetization process starts from zero instantly after imposing the staggered field and boosts monotonically up to attain a saturated state. It is explicit that almost the same growing behaviour also takes place for when the system is located in the gapped $C1$ and $C2$ phases [Figs.~{\color{blue}\ref{St-Mag-Lanc}(c)} and {\color{blue}\ref{St-Mag-Lanc}(d)}].


\section{Summary} 
Quantum phase transition originated from quantum fluctuations at absolute zero temperature deals with dramatic
changes of the ground-state and low-excitation properties. Despite the many-body systems, it has been  found that it may also outstretch in a few-body system. Understanding emergent quantum phenomena is a major challenge of physics and a requirement for future technologies that aim to manipulate in a controlled way the quantum properties of many-spin systems.

In this paper, we have studied the ground-state magnetic phase diagram of a novel quantum spin chain, a 1D spin-1/2 XXZ model in the presence of the modulated period of two lattice units Dzyaloshinskii-Moriya interaction (DMI) and alternating with the same period longitudinal magnetic field. In this respect, we first have focused on the exactly solvable case of the Hamiltonian where $J_z=0.0$. In the absence of the magnetic fields, the ground state is constructed from a gapped composite phase made of the coexistence of a long-range dimer and an alternating spin chirality.
The driven outcomes have indicated that the uniform magnetic field generates two QPTs.  The gapped mentioned composite phase endures in the presence of a uniform  field up to a critical field. Above endurance is emerged as a zero-plateau in the magnetization curve. More increasing the uniform field, a gapless phase is appeared up to the second critical field, where the magnetization is saturated.  On the other hand, instantly a staggered magnetic field is exerted, the Neel LRO creates and coexists with dimer and alternating spin chirality orders.

In order to consider the non-integrable Hamiltonian ($J_z \ne 0$), we have employed two apparatuses; the continuum-limit bosonization approach as an analytical technique and the Lanczos algorithm as a numerical method to confirm the analytical results.
Our results in this condition have unveiled that the uniform field behaves the same as what has come out for the case $J_z=0.0$ with this difference that the critical fields in addition to DMI, now, are dependent on the value of $\gamma$.
The staggered magnetic field manifests three different behaviors in the range of $\gamma$. When the system is settled in the FM phase, it does not change the phase up to a critical staggered field where after that, it induces the Neel phase in the system. When the system is put in the LL phase, as soon as it is exerted, the Neel phase is created. Eventually, in the case where the system is placed in the composite $C1$ ($C2$) phase, the staggered field just creates (amplifies) the Neel phase.


\begin{acknowledgements}

G.I.J. acknowledges support from the Shota Rustaveli Georgian National Science Foundation through the grant N FR-19-11872.
In addition, S. Mahdavifar and H. Cheraghi wish to acknowledge the support of the Iran National Science Foundation
(INSF) under grant number 98018317.
\end{acknowledgements}

\section*{Appendix:  Calculation of the order parameters }
The calculations of the alternating chirality and dimer order parameters are straightforward.
Applying ({\color{blue}\ref{eq3}}) to original definition of these parameters and afterwards using the Bogoliobov transformation, and also  ({\color{blue}\ref{eq24}}) and ({\color{blue}\ref{eq25}}), lead to
\begin{eqnarray}
{\cal D}^{\perp} &=& \frac{1}{N}\sum\limits_{n = 1}^{N} (-1)^{n} {\left\langle S_n^x S_{n+1}^x+ S_n^y S_{n+1}^y\right\rangle }   \nonumber\\
 &=&\frac{1}{2N}\sum\limits_{n = 1}^{N/2} {\left\langle { - a_n^\dag {b_n} - b_n^\dag {a_n} + a_{n + 1}^\dag {b_n} + b_n^\dag {a_{n + 1}}} \right\rangle }  \nonumber\\
 &=& -\frac{ 1}{2N}\sum\limits_{k } {\sin (2{\Phi _k})\omega ({\theta _k},k)} \left\langle {\alpha _k^\dag {\alpha _k} - \beta _k^\dag {\beta _k}} \right\rangle  \nonumber\\
& =&\frac{1}{2N}\sum\limits_{k\rq{}} \sin(2\phi_{k\rq{}}) \omega (\theta _{k\rq{}},k\rq{})\,  
\end{eqnarray}
and
\begin{eqnarray}
{\cal K}^z &=& \frac{1}{N}\sum\limits_{n = 1}^{N} (-1)^{n} {\left\langle S_n^x S_{n+1}^y- S_n^y S_{n+1}^x\right\rangle } \nonumber\\
&=&\frac{{ - i}}{{2N}}\sum\limits_{n = 1}^{N/2} {\left\langle {a_n^\dag {b_n} - b_n^\dag {a_n} - b_n^\dag {a_{n + 1}} + a_{n + 1}^\dag {b_n}} \right\rangle }   \nonumber\\
&=&  \frac{1}{{2N}}\sum\limits_{k } {\sin (2{\Phi _k})\varpi ({\theta _k},k)} \left\langle {\alpha _k^\dag {\alpha _k} - \beta _k^\dag {\beta _k}} \right\rangle  \nonumber\\
& =&\frac{1}{2N}\sum\limits_{k\rq{}} \sin(2\phi _{k\rq{}}) \varpi (\theta _{k\rq{}},k\rq{})\,.  
\end{eqnarray}
where as was mentioned before,  $k\rq{}$ refers to $[-\pi/2,\pi/2]$ for $H<H_{c_1}$ and $\Lambda (k)$ for  $H>H_{c_1}$.

\bigskip{}

\vspace{0.3cm}

\end{document}